\documentclass[twocolumn]{aastex631}

\usepackage{graphicx}
\usepackage[caption=false]{subfig}
\usepackage{amsmath}
\usepackage{appendix}
\usepackage{xspace} 
\newcommand{\Rvir}{$R_{500}$\xspace}
\newcommand{\Mvir}{$M_{500}$\xspace}
\newcommand{\Planck}{\textit{Planck}\xspace}
\newcommand{\WMAP}{\textit{WMAP}\xspace}
\newcommand{\Msun}{$\mathrm{M_{\odot}}$\xspace}
\newcommand{\ymap}{$y$-map\xspace}
\newcommand{\ymaps}{$y$-maps\xspace}
\newcommand{\kms}{$\mathrm{km}~\mathrm{s}^{-1}$\xspace}

\defcitealias{KTully17}{KT17}
\defcitealias{Arnaud10}{A10}
\defcitealias{LeBrun15}{L15}

\bibpunct[; ]{(}{)}{,}{a}{}{;}

\graphicspath{{./}{figures/}}

\shorttitle{Resolved Groups}
\shortauthors{Pratt, Qu, and Bregman}

\begin{document}

\title{The Resolved Sunyaev\textendash Zel'dovich Profiles of Nearby Galaxy Groups}

\author{Cameron T. Pratt}
\author{Zhijie Qu}
\author{Joel N. Bregman}
\affil{Department of Astronomy, University of Michigan, Ann Arbor, MI 48104}



\begin{abstract}

Much of the baryons in galaxy groups are thought to have been driven out to large distances ($\gtrsim$\Rvir) by feedback, but there are few constraining observations of this extended gas. This work presents the resolved Sunyaev\textendash Zel'dovich (SZ) profiles for a stacked sample of 10 nearby galaxy groups within the mass range log$_{10}(M_{500}[M_{\odot}]) = 13.6 -13.9$. We measured the SZ profiles using the publicly available \ymap from the Planck Collaboration as well as our own \ymaps constructed from more recent versions of \Planck data. The \ymap extracted from the latest data release yielded a significant SZ detection out to 3 \Rvir. In addition, the stacked profile from these data was consistent with simulations that included AGN feedback. Our best-fit model using the latest \Planck data suggested a baryon fraction $\approx 5.6\%$ within \Rvir. This is significantly lower than the cosmic value of $\approx 16\%$, supporting the idea that baryons have been driven to large radii by AGN feedback. Lastly, we discovered a significant ($\sim 3\sigma$) ``bump'' feature near $\sim 2$ \Rvir that is most likely the signature of internal accretion shocks.

\end{abstract}


\section{Introduction}
In the present epoch, most of the baryons in the Universe have fallen into deep potential wells carved out by galaxies, galaxy groups, and galaxy clusters. Strong gravitational forces from the most massive dark matter systems, namely galaxy groups and clusters, have condensed much of the accreted material into hot, gaseous halos known as the intragroup medium (IGrM) and intracluster medium (ICM). One might expect these halos to heat in a self-similar manner since gravity is a scale-free force \citep{Kaiser86}. In reality, this self-similarity scaling is perturbed by non-gravitational processes such as stellar feedback, active galactic nucleus (AGN) feedback, radiative cooling, and accretion shocks. 

Observational and theoretical work has provided strong evidence that AGN feedback is likely the main culprit in disrupting the atmospheres of galaxy groups and clusters \citep[see][for a review]{Gitti11}. AGN jets blast energy into the ICM/IGrM, expanding to tens of kpc, and carve out large cavities in the X-ray emitting gas \citep{Gaspari13,Liu19,Lacy19,Pasini20}. These outflows heat the surrounding medium and drive low entropy gas away from the central regions. This idea gained attention when trying to explain the observed lack of star formation in the center of galaxy clusters, known as the ``cooling flow problem'' \citep{McNamaraCoolingFlow2007}. 

Even in the presence of AGN feedback, gas is still able to cool and condense inside the halo, especially near the edges of the X-ray bubbles. The cold gas then rains back down onto the galaxies, refuels the central engine, and induces a self-regulated cycle \citep{Gaspari11,Gaspari13,Li15, Gaspari17,McDonald18,Tremblay18,McDonald19,Russell19}. Nevertheless, there is still debate regarding which physical mechanisms are most responsible for regulating these outflows \citep{Gaspari20}. Many parameters still need to be tuned by theorists in order to match observations such as: the accretion zone size, the amount of kinetic and thermal feedback, the opening jet angle, ambient gas density, and the angular momentum of the accreting gas \citep{Meece17,Prasad20}. 

To further complicate this picture, the effects of feedback are predicted to depend on the halo mass. Given the fact groups exist in shallower potential wells than clusters, outflows from galaxy groups do not have to fight as hard to propagate outward. Therefore, AGN outflows may be capable of driving gas from the central regions of galaxy groups out to large distances \citep[$\gtrsim$ \Rvir\footnote{\Rvir is the radius at which the density is 500$\times$ the critical density of the Universe.};][]{Roychowdhury05,Mukherjee19,Ishibashi16}. In turn, one would expect to see a deficit of gas within \Rvir relative to the self-similar prediction for low-mass systems, and this prompted the search for the predicted ``breaking'' of scaling relations \citep{Eckmiller11,Gaspari14,Lovisari15,Paul17,Farahi18,Pratt20, Sereno20}. Much of the observational work to-date has focused on the gas properties within \Rvir, but very few measurements have been made of the extended gas.

X-ray observations have given excellent insight on the thermal profiles of the inner parts of the IGrM and ICM due to the squarded dependence of the density in the emission measure. On the other hand, the Sunyaev\textendash Zel'dovich (SZ) effect is linearly dependent on the gas density which makes it a more sensitive tracer of diffuse gas at large radii. Both X-ray and SZ observations have been used to study massive galaxy clusters ($M_{500}\gtrsim 10^{14} M_{\odot}$), revealing that cluster thermal profiles appear remarkably self-similar outside their cores \citep{Pratt09,Arnaud10,Sayers13, Mantz16}. Galaxy groups ($M_{500}\lesssim 10^{14} M_{\odot}$), however, are not expected to show the same degree of self-similarity as rich clusters. In particular, galaxy groups are expected to exhibit flatter pressure profiles caused by AGN feedback \citep[][hereafter \citetalias{LeBrun15}]{LeBrun15}. 

The impact of AGN feedback on the pressure profiles of galaxy groups can be tested with resolved SZ signals. Currently, the best publicly available all-sky SZ maps (\ymaps) were constructed from \Planck data using two modified {\it{Internal Linear Combination}} (ILC) algorithms \citep{PlanckYMAP}. They were generated using the second version of \Planck data (PR2) and were labeled as the \textit{NILC} and \textit{MILCA} \ymaps. These \ymaps have led to substantial studies of unresolved, massive galaxy clusters. Herein, we refer to the \textit{NILC} map as the ``public \ymap'' and eventually call it ``PR2NILC''; we do not report on \textit{MILCA}.

Only nearby objects are resolved by the coarse resolution of the public \ymap (full width at half maximum [FWHM] = 10{\arcmin}). Resolved SZ profiles have been measured for nearby clusters \citep{PlanckClusterProfile,PlanckComa,PlanckVirgo}, but little attention has been given to nearby galaxy groups. In this work, we investigate the resolved SZ profiles of 10 galaxy groups within 3500 \kms from the catalog produced by \citep[][hereafter \citetalias{KTully17}]{KTully17}.   

The rest of the paper is structured in the following way: 
\autoref{sec:sample} describes the sample selection of galaxy groups from the \citetalias{KTully17} catalog; \autoref{sec:NILC} presents the \textit{Needlet Internal Linear Combination} (NILC) method used to construct our own \ymaps from newer and improved versions of \Planck data; \autoref{sec:extraction} presents the methodology used to extract the radial SZ profiles, remove sources of contamination, calculate uncertainties, and estimate bias corrections from Monte Carlo simulations; \autoref{sec:results} presents the results of the stacked SZ profile; \autoref{sec:discussion} discusses the differences between \ymaps, presents estimates of the baryon fraction, and addresses the extended part of the SZ profile; and \autoref{sec:summary} provides a brief summary of our work. Throughout this work we assume a cosmology with a Hubble constant of $H_{0} = 70 ~\mathrm{km~s^{-1}~Mpc^{-1}}$, and density parameters of $\Omega_{m} = 0.3$ and $\Omega_{\Lambda} = 0.7$.
\section{Sample of Galaxy Groups}\label{sec:sample}

The sample of galaxy groups were drawn from the catalog constructed by \citetalias{KTully17}. It was assembled mainly using the 2MASS Redshift Survey, although it included archival data of low surface brightness systems to supplement their grouping algorithm. They limited their group sample to systems with recessional velocities  $<$3500 \kms (Hubble distance $\approx$ 50 Mpc). At this distance, the survey was complete to $L^{*}$ in the $K_{s}$ band. 

All galaxies had measured radial velocities, and an additional 19\% also had independent distance estimates from \textit{Cosmicflows-3} \citep{CosmicFlows3}. \textit{Cosmicflows-3} is a compilation of galaxy distances that has been built over the last decade \citep{Tully08,Tully13} and is the latest version available. It is particularly useful to have independent distance measurements for galaxies in the nearby Universe since the magnitude of their peculiar motions may be comparable to their recessional velocities. All 10 objects in our final sample had reliable distance estimates ($\leq 15 \%$ error), although this was not true for every system in the \citetalias{KTully17} catalog.

The distance estimates were then combined with the apparent magnitudes in the $K_{s}$ band to estimate luminosities. The $K_{s}$-band luminosity was then used as a proxy for the stellar mass. Stellar masses were converted into halo virial masses, $M_{v}$, using scaling relations from \citet{Tb15}. It is more customary, however, to work in units of $M_\Delta$, where $M_{\Delta}$ is the mass enclosed by the radius defined by the overdensity parameter, $\Delta$, relative to the critical density. Throughout this work we use $\Delta = 500$. In order to compare these masses with other works, we converted $M_{v}$ to $M_{500}$ using the relation $r_{200} = 0.198 (\frac{M_{v}}{10^{12} M_{\odot}})^{1/3}h_{75}^{-2/3}$ from \citet{Tully15} and assuming $r_{500} \approx 0.7r_{200}$.

For the purposes of this study, the \citetalias{KTully17} catalog was filtered to select resolved systems that were fairly isolated on the sky. The criteria for our selection included:
\begin{enumerate}
    \item $10^{13.6} M_{\odot} < M_{500} <  10^{14} M_{\odot}$.
    
    \item  The angular size of \Rvir was $> 40\arcmin$ to ensure the gas profiles were resolved in the \ymaps, i.e., 0.25 \Rvir $\geq$ 10{\arcmin}.
    
    \item Galactic latitudes were $|{\it{b}}| > 15^{\circ}$ to avoid the most extreme contamination from the Galactic plane. We also masked regions where the SZ signal was highly correlated with the dust emission for each localized field (see \autoref{sec:contamination}).
    
    \item The number of member galaxies was $\geq 3$.
    
    \item Systems were required to possess a redshift-independent distance measurement.
    
    \item Center positions of the objects were not within \Rvir or $20{\arcmin}$ (whichever was larger) of another known galaxy group/cluster from the \textit{meta-catalogue of X-ray detected clusters of galaxies} \citep[MCXC;][]{MCXC}, the \textit{second Planck catalogue of Sunyaev\textendash Zel'dovich sources} \citep[PSZ2;][]{PlanckSZCatalog}, or the \citetalias{KTully17} catalog. For this procedure, we used objects with \Mvir $>10^{12} M_{\odot}$ from the \citetalias{KTully17} catalog. This ensured the objects of interest were isolated on the sky from other known groups/clusters.
\end{enumerate}

\begin{figure}[htbp]
   \includegraphics[width=0.45\textwidth]{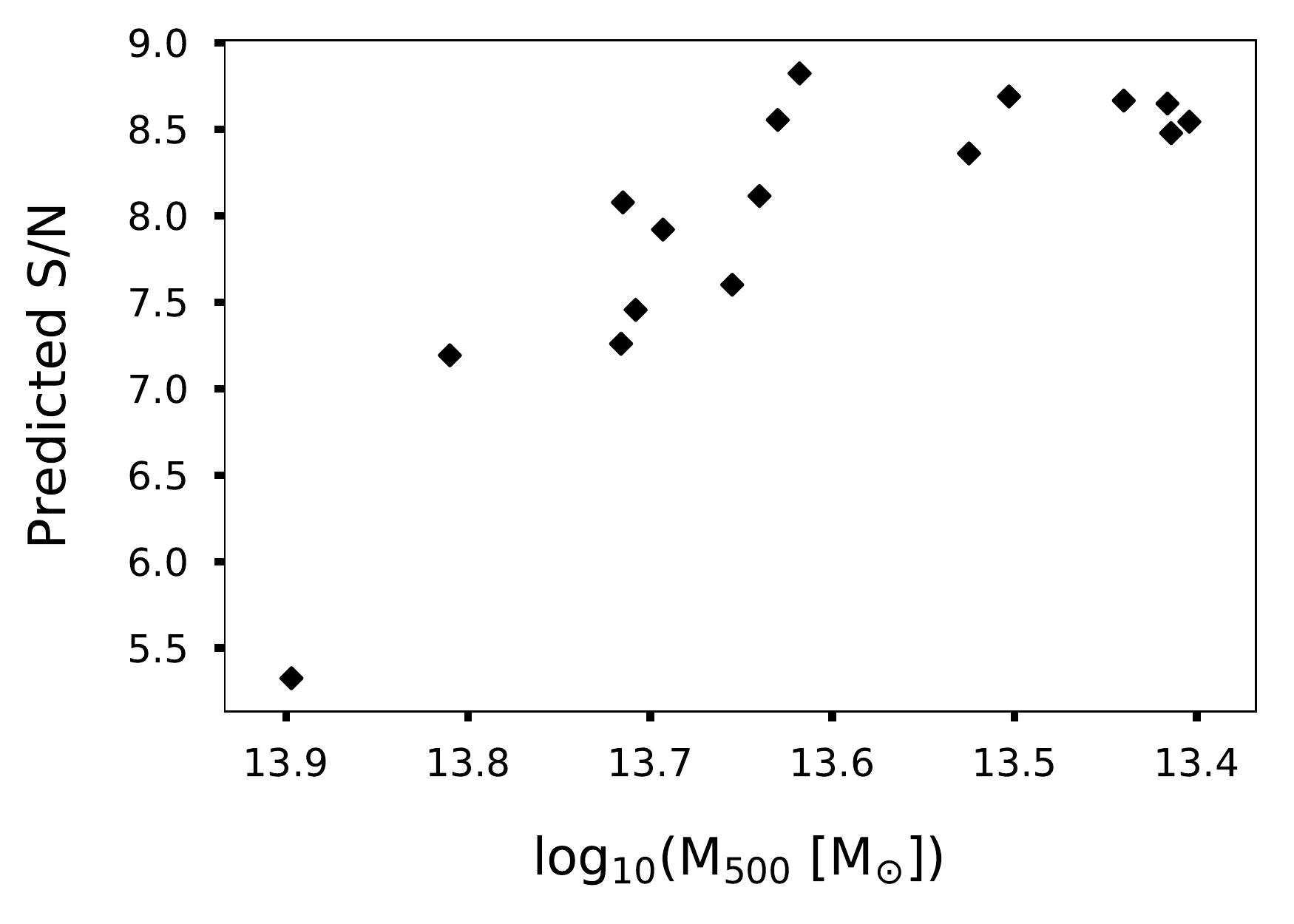}
    \caption{The predicted S/N of stacked groups by subsequently adding lower mass objects to the sample (note the backwards mass scale). For clarity, the second leftmost datum is the S/N one would expect when stacking the two most massive objects in the sample. The stacked S/N appears to stop increasing around \Mvir $\sim 10^{13.6}$ \Msun and we used this as a threshold for the lower mass limit.}%
    \label{fig:snr_predict}%
\end{figure}

The upper mass limit was selected because lower mass objects have not received a lot of attention in the literature, and all of the more massive objects in the catalog were well-known clusters (e.g., Virgo and Centaurus). We also imposed a lower mass limit for stacking purposes. One could consider halos within a mass range $\sim$ 0.5 dex to be similar and stack them together. However, \autoref{fig:snr_predict} shows the predicted signal-to-noise ratio (S/N) eventually stopped increasing at a certain mass. This was determined by calculating the predicted signal for each object with \Mvir $< 10^{14}$ M$_{\odot}$ and assessing the noise levels in the PR2NILC \ymap. The signal was estimated by finding the average SZ value within 0.25 \Rvir after convolving the AGN 8.0 pressure profile from \citetalias{LeBrun15} with the point spread function (PSF), assumed to be a 2D Gaussian beam of FWHM = {10\arcmin} \citep{PlanckYMAP}. The associated noise was estimated using the method described in \autoref{sec:uncertainty}. Starting with the most massive object, we subsequently included lower mass systems and stacked their signals. The S/N of the stack stopped increasing once objects below \Mvir $< 10^{13.6}$ M$_{\odot}$ were included, therefore, we did not consider systems below this mass.

\startlongtable
\begin{deluxetable*}{rrrlcccccl}
\tabletypesize{\footnotesize}
\tablecaption{Sample of Galaxy Groups \label{tab:sample_info}}
\tablehead{\colhead{PCG} & \colhead{$l$} & \colhead{$b$} & \colhead{MCXC Name} & \colhead{log$_{10}$($M_{500}$)} & \colhead{$R_{500}$} & \colhead{Velocity} & \colhead{Distance} & \colhead{Distance \% Error} &  \colhead{BCG Name} \\\colhead{} & \colhead{$[^{\circ}]$} & \colhead{$[^{\circ}]$} & \colhead{} & \colhead{[log$_{10}$([$M_{\odot}$])]}& \colhead{[arcmin]} & \colhead{[km~s$^{-1}$]} & \colhead{[Mpc]}& \colhead{[\%]} & \colhead{}}
\startdata
39659 & 281.80280 & 67.37240 &                                                             & 13.90 & 71.9 &  2224    & 31.4 & 4 &  NGC4261       \\
13418 & 236.71250 & -53.63970 &  MCXC J0338.4-3526                                          & 13.81 & 116.4 &  1502    & 18.1 & 3 &  NGC1399       \\
64192 & 350.90020 & -32.63890 &  MCXC J2009.9-4823                                          & 13.72 & 58.0 &  2916    & 33.9 & 4 &  NGC6868       \\
49356 & 82.61060 & 71.63340 &                                                             & 13.71 & 61.2 &  2408    & 32.0 & 6 &  NGC5353       \\
45174 & 306.20280 & 32.27490 &  MCXC J1304.2-3030                                          & 13.71 & 44.4 &  3292    & 44.0 & 8 &  NGC4936       \\
46247 & 311.23340 & 46.09540 &  MCXC J1315.3-1623                                          & 13.69 & 69.2 &  2488    & 27.9 & 5 &  NGC5054       \\
40498 & 297.48480 & 22.83550 &                                                             & 13.65 & 65.8 &  3258    & 28.5 & 6 &  NGC4373       \\
33871 & 281.57860 & 21.08900 &                                                             & 13.64 & 60.0 &  2864    & 30.9 & 6 &  NGC3557       \\
53932 & 0.42740 & 48.79460 &  MCXC J1506.4+0136                                          & 13.63 & 68.7 &  1819    & 26.7 & 4 &  NGC5846       \\
62870 & 339.69610 & -25.13320 &                                                             & 13.62 & 60.6 &  3443    & 30.0 & 15 &  NGC6753   \\
\enddata
\tablecomments{The first column (1) is the ``PCG'' identification number used to distinguish galaxy groups in \citetalias{KTully17}; (2) and (3) are the galactic coordinates in degrees; (4) is the identification of the system in the MCXC catalog if it existed; (5) is the logarithm of the estimated mass (in solar units) within \Rvir; (6) is the angular size of \Rvir for each system in arcminutes; (7) is the recessional velocity of each galaxy group; (8) is the measured distance and (9) is its fractional uncertainty; (10) is the NGC identification for the BCG.}

\end{deluxetable*}

In the \citetalias{KTully17} catalog, the luminosity-weighted positions of the galaxies were denoted as the center of each system. This position, however, was not the ideal center to extract the SZ signal. It is known that the center of the halo gas typically shares the same position of the brightest cluster galaxy \citep[BCG;][]{George12}. We defined the BCG as the most luminous galaxy within 0.15 \Rvir of the luminosity-weighted center if it existed, otherwise, we considered the most luminous galaxy within \Rvir to be the BCG. We searched within 0.15 \Rvir first in case there was a brighter galaxy a little further away from the luminosity-weighted center; in other words, preference was given to bright galaxies close to the luminosity-weighted center. Indeed, we found that 4/5 (80\%) of the systems with X-ray information from the MCXC catalog were located closer to the BCG than the luminosity-weighted center. In cases where groups had center positions from MCXC, we adopted the X-ray position to be the center of the SZ extraction, otherwise, we used the position of the BCG. All of the relevant information on these systems is presented in table \autoref{tab:sample_info}. 
\section{SZ Map Making}\label{sec:NILC}

Here we present the data and methods used to create all-sky SZ maps (\ymaps). In particular, we describe our use of data from the \Planck mission and the \textit{Wilkonson Microwave Anisotropy Probe} (\WMAP). Then we present our NILC method to create \ymaps.

\subsection{Frequency Maps}

The public \ymap from \citet{PlanckYMAP} was created using the second version of \Planck data \citep[][``PR2'']{PlanckPR2}, however, there have been significant improvements to the \Planck data processing in more recent years. For example, the third release of \Planck data \citep[][``PR3'']{PlanckPR3} improved the \textit{Low Frequency Instrument} pipeline by including Galactic emission and the CMB dipole in their calibration. Also, the \textit{High Frequency Instrument} processing omitted 22 days of data taken during periods of increased Solar activity. Finally, PR3 implemented a different map-making algorithm that used foreground templates from PR2 to reduce systematic effects.

Further advancements were made in the fourth release of \Planck data \citep[][``PR4'']{PlanckPR4}. For instance, PR4 increased the total integration time by about $8\%$ relative to PR3 by including data taken during the repointing periods of the satellite. In addition, the  data processing of PR4 identified and removed the artificial ringing patterns in the 353 GHz map from PR3 called ``zebra stripes''. PR4 also made improvements in suppressing another type of striping feature from turning time-ordered data of the sky into a single image. These stripes occur along the scanning pattern of the satellite and can produce significant amplitudes if not dealt with properly \citep{Ashdown07}. Various ``destriping'' algorithms have been applied throughout the \Planck data releases, but PR4 implemented an extremely short baseline step of 167 ms. This short baseline reduced the noise levels on scales $\lesssim 5^{\circ}$ relative to the larger baselines used in PR3 and PR2. Finally, PR4 yielded less systematic noise, on all angular scales, compared to previous releases. 

The public \ymap was made exclusively with \Planck data. This \ymap exhibited strong residuals from radio sources due to the limited information at low frequencies. Radio sources with a power law emission will create negative ``holes'' in the \ymap. This is because they cause an excess of power at low frequencies whereas the SZ effect causes a decrement. For our SZ extraction, we included more low frequency coverage with all-sky data from \WMAP to better handle radio source contamination. Specifically, we used the 9-year \WMAP data which provided an additional five bands that covered the 23 - 94 GHz range \citep{WMAP9yr}.

Working with heterogeneous data sets (\Planck and \WMAP) came with notable caveats. PR4 data retained the Solar dipole in all frequency maps, but this component was deprojected in PR2, PR3 and \WMAP data. Moreover, the CMB calibrations were all different among these data sets. For these reasons, we performed the SZ extraction using internal difference maps\textemdash that is the difference between subsequent frequency maps for a given data set. Taking the difference maps effectively nullified the CMB and solar dipole components, allowing us to work with PR4/PR3 and \WMAP data together. In the next section we describe our extraction of the SZ signal using these internal difference maps. Moving forward, our \ymaps created with the difference maps from PR4/PR3 and \WMAP data will be labeled as ``PR4NILC'' and ``PR3NILC'' respectively, while the \textit{NILC} \ymap from \citet{PlanckYMAP} will be referred to as ``PR2NILC''.

\subsection{Internal Linear Combination}

The SZ effect measures the energy boost gained by \textit{Cosmic Microwave Background} (CMB) photons after they interact with hot electrons in the ICM or IGrM. It causes a decrease in the flux relative to the CMB at frequencies $<$ 217 GHz and an increase at higher energies. The SZ surface brightness is measured as a dimensionless quantity, $y$, which is given by
\begin{equation}\label{eq:sz_temp}
    \frac{\Delta T}{T_{\mathrm{CMB}}} = f(\nu)y
\end{equation}
where $f(\nu)$ is the SZ spectral dependence, $\frac{\Delta T}{T_{\mathrm{CMB}}}$ is the change in temperature relative to the CMB. The formal analytical expression for $f(\nu)$ is
\begin{equation}
    f(x) = \bigg( x \frac{e^{x}+1}{e^{x}-1} - 4\bigg)(1+\delta (x,T_{\mathrm{e}}))
\end{equation}
where $x = \frac{h\nu}{k_{\mathrm{B}}T_{\mathrm{CMB}}}$ ($k_{\mathrm{B}}$ is the Boltzmann constant, $h$ is Planck's constant, and $T_{\mathrm{CMB}}$ is the temperature of the CMB), $T_{\mathrm{e}}$ is the temperature of the electrons, and $\delta$ is a relativistic correction \citep{Itoh98,Carlstrom02}. Relativistic corrections must be applied to extremely high-mass systems (\Mvir $\sim 10^{15}$ \Msun) but are negligible for galaxy groups. Also, one must calculate the effective $f(\nu)$ for wide-band observations by integrating over the instrumental transmission window.     

The SZ signal is just one of many astrophysical components that contributes to the observed intensity. As mentioned above, using difference maps (in thermodynamic units [$K_{\mathrm{CMB}}$]) effectively cancels the contributions from the CMB and the Solar dipole. An internal difference map between two frequency channels, $X_{\Delta}$, can be expressed with a single component model
\begin{equation}\label{eq:ILC}
    X_{\Delta}(p) = f_{\Delta}S_{SZ}(p) + N_{\Delta}(p)
\end{equation}
where $f_{\Delta}$ is the SZ spectral dependence of the difference map, $S_{SZ}$ is the SZ signal, and $N_{\Delta}$ is the noise term which is usually dominated by foreground Galactic emission. For this single component model, the ILC estimator of the SZ signal \citep{Delabrouille09} is given by \begin{equation}\label{eq:ILC_estimator}
    \widehat{S_{SZ}}(p) = \frac{\mathbf{f}_{\Delta}^{t}~(\mathbf{\widehat{R}}_{\Delta}(p))^{-1}}{\mathbf{f}_{\Delta}^{t}~(\mathbf{\widehat{R}}_{\Delta}(p))^{-1}~\mathbf{f}_{\Delta}} ~\mathbf{x}_{\Delta}(p)
\end{equation}
where $\mathbf{f}_{\Delta}$ is the vector of the differences in the SZ frequency dependence, $\mathbf{x}_{\Delta}$ is the vector of difference maps, and $\widehat{\mathbf{R}}$ is the empirical covariance matrix of the difference maps \citep{Erikson04}. In the standard ILC method, the covariance matrix is calculated using the entire sky.

In practice, it is difficult to model all of the different foreground noise components. Without adequate models for all components, there will be inevitable contamination, and this is one of the shortcomings of the standard ILC method. This motivated the work to construct new ILC methods that could help reduce the effects of contamination, most commonly by localizing the algorithms in harmonic space, pixel space, or both. In this work, we focus on using both harmonic and spatial localizations through the application of spherical needlets.

\subsection{Needlet Analysis}
A popular technique used to reduce the effects of contamination is the NILC method. Historically, NILC algorithms have been used to extract the CMB as well as the SZ effect \citep{Delabrouille09, Remazeilles11,Hurier13,Remazeilles13,PlanckYMAP}. The basic idea of the NILC algorithms is to first filter the all-sky maps in harmonic space and then perform the SZ extraction using spatially localized regions. The key advantages of the NILC method are: (1) allowing one to combine intensity maps of different resolutions and from different instruments (2) reducing the noise properties over various angular scales. Below we describe the properties of the needlets used in this work and also lay out the steps for the NILC algorithm.

For the NILC method, one must define the shapes for a set of multipole filters along with a local extraction region to calculate the covariance in \autoref{eq:ILC_estimator}. The covariance between 2 difference maps ($a$ and $b$) is given by 
\begin{equation}
\label{eq:covariance}
    \mathbf{\widehat{R}}_{ab}^{(j)}(p) = \frac{1}{N_{p}} \sum_{p^{'}\in \mathcal{D}_{p}} x_{a}^{(j)}(p^{'})x_{b}^{(j)}(p^{'})w^{(j)}(d)
\end{equation} where $N_{p}$ is the number of needlet modes for a given harmonic scale, $j$, $\mathcal{D}_{p}$ is the pixel domain used to compute the covariance, and $w$ is Gaussian weighting that depends on the angular separation between pixels, $d$. Notably, the covariance was calculated on the data themselves. This is important as there exists random correlations between the SZ signal and the local contamination. Minimizing the local variance with \autoref{eq:ILC_estimator} inevitably suppresses the SZ signal\textemdash an effect known as the ILC bias \citep{Delabrouille09,Remazeilles13}. 

The ILC bias is regulated by $N_{p}$ as a result of the multipole window function and pixel domain, where a smaller value of $N_{p}$ produces a larger bias. Moreover, $N_{p}$ also determines the residual variance, yielding less uncertainty in the \ymap for a smaller $N_{p}$. Thus, there is a trade-off between lowering the bias and raising the variance (or vice versa) that is specific to each set of needlets. This quantity, however, was not stated explicitly by \citet{PlanckYMAP}, and this prevented us from investigating the amount of bias in PR2NILC. In \autoref{sec:bias} we estimate the bias for our set of needlets and show the correction is small ($\lesssim 5\%$) for the sample of galaxy groups. 

\begin{figure}[t!]
   \includegraphics[width=0.45\textwidth]{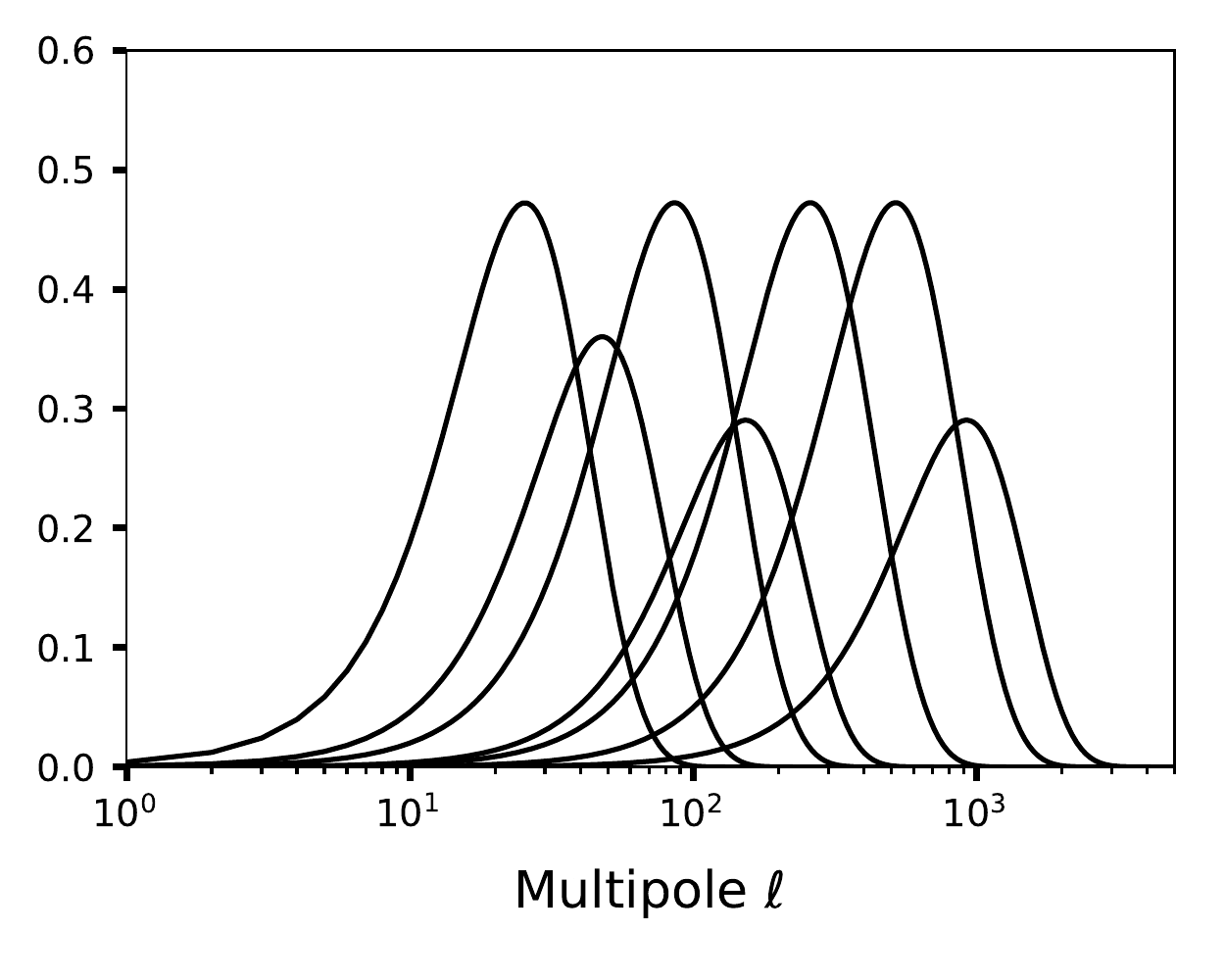}
    \caption{Needlet window functions used to filter the difference maps in multipole space. These filters captured features on scales $\gtrsim$ 10\arcmin and $\lesssim 10^{o}$}.%
    \label{fig:needlet}%
\end{figure}

Here we explicitly state the shapes and sizes of the needlets used in our algorithm. The shapes of the needlets consisted of 7 Gaussian beam window functions (indexed by $j$) in harmonic space, $h^{(j)}$. Each window was characterized by the Gaussian FWHMs: 10, 15, 30, 60, 90, 180, 300, and 600 arcminutes. We subtracted the subsequent Gaussian functions to obtain the windows: 10{\arcmin}-15{\arcmin}, 15{\arcmin}-30{\arcmin}, {30{\arcmin}-60{\arcmin}} etc... They are presented as a function of the multipole moment, $\ell$, in \autoref{fig:needlet}. 

We spatially localized the SZ extraction for a single pixel using the information from surrounding pixels. This included setting the $\sigma$ of the Gaussian weights in \autoref{eq:covariance} to be 10$\times$ the upper bound of each window function called FWHM$_{\mathrm{up}}$. The surrounding pixel domain was defined out to a radius of $30~\times$ FWHM$_{\mathrm{up}}$. Using the 10{\arcmin}-15{\arcmin} window as an example, FWHM$_{\mathrm{up}}$ = 15{\arcmin} and the width of the Gaussian used for the weights was $\sigma = 10~\times$ FWHM$_{\mathrm{up}} = 150{\arcmin}$. 

Now we describe the NILC procedure which closely followed the discussions in \citet{Delabrouille09} and \citet{Remazeilles13} to which we refer the reader to for more detailed information. Specifically, this follows the \textit{analysis} and \textit{synthesis} steps laid out by \citet[][and references therein]{Delabrouille09}.
\begin{itemize}
    \item First, we performed the spherical harmonic transformation on each of the internal difference maps. Then we applied the first half of the window function ($\sqrt{h^{(j)}}$) to filter the Fourier coefficients in multipole space. Due to the different instrumental resolutions between the frequency maps, only some of the difference maps were considered for each window function. It was required that the instrumental resolution of the lowest frequency used in the difference map was less (better) than FWHM$_{\mathrm{up}}$. As an example, consider the {15\arcmin-30\arcmin} window function. For this window, the lowest frequency used in the difference maps had a resolution that was better than {30\arcmin} in order to be included in the extraction. The lowest frequency difference map used for \Planck data was the 44-70 GHz difference map \citep[FWHM = {28.09\arcmin} and {13.08\arcmin} respectively;][]{PlanckOverview}. Moreover, the 30-44 GHz difference map was not included since the resolution of the 30 GHz map \citep[FWHM = {33.16\arcmin};][]{PlanckOverview} was worse than {30\arcmin}.
    
    \item Next we performed the ILC component separation using \autoref{eq:ILC_estimator} on the filtered maps. An intermediate SZ map was obtained after adding all maps of different scales together.
    
    \item The last step was to perform the spherical harmonic transformation on the intermediate SZ map and apply the second half of the window functions in harmonic space. Applying the inverse transform of these filtered Fourier coefficients yielding the final \ymap.
\end{itemize}
\section{SZ Extraction of Galaxy Groups}\label{sec:extraction}

In this section, we describe our procedure to extract the SZ signal from different \ymaps. First, we explain how we identified and excluded strong sources of contamination. Second, we describe our estimations of the uncertainty in localized fields. Third, we present two methods for stacking the galaxy groups together. Finally, we estimate and correct for the bias with Monte Carlo simulations.
\subsection{Contamination}\label{sec:contamination}
\subsubsection{Small-Scale}

Background fluctuations in the \ymaps occurred both in the positive and negative directions. The power spectrum at large multipole moments (small angular scales) was dominated by galaxy clusters (positive), infrared (IR) sources (positive), and radio sources (negative). We investigated the effects of radio and IR sources using those identified in the second Planck Catalogue of Compact Sources \citep[PCCS;][]{PCCS}. 

The analysis of radio and IR sources started by generating 20{\arcmin}$\times$ 20{\arcmin} pixel images of PR2NILC data. Each image contained a field-of-view $15^{\circ} \times 15^{\circ}$ and was centered around the objects in \autoref{tab:sample_info}. Next, we smoothed the maps using a 2D Gaussian filter with a FWHM $= 5^{\circ}$ and subtracted them from the raw maps in attempt to remove large-scale fluctuations. Passing the maps through this high-pass filter allowed us to analyze the amount of small-scale contamination. Specifically, we focused on the pixels within the upper/lower 1\% in each field. Throughout this analysis, we excluded pixels associated with known galaxy clusters identified in the MCXC \citep{MCXC} and PSZ2 \citep{PSZ2} catalogs since they produced strong positive signals. 

The extreme negative pixels (lower 1\%) were most associated with PCCS sources in the 143 GHz channel. About 25\% of the extreme negative pixels were near a radio source detected in the 143 GHz band, with fluxes typically above 460 mJy. The extreme positive pixels (upper 1\%), however, were not as strongly correlated with the IR sources; only $\approx$ 5\% of the extreme positive pixels shared positions with 857 GHz PCCS sources with fluxes above 1130 mJy. This was not surprising given the weak SZ spectral dependence at high frequencies, hence, IR sources did not severely propagate into the \ymaps. 

While radio and IR sources account for some of the extreme positive and negative residuals, they did not account for the majority of the extreme valued pixels. The \ymap pixel distributions possessed long-tailed wings, especially in the negative direction, and most of them were not identified as radio sources. Instead, these extreme regions were likely artificial features introduced by the NILC method. We considered these regions to be a source of contamination, so we excluded such regions out to a radius of 20{\arcmin} as an effort to produce a flatter background. The same exclusion procedure was applied to the radio/IR sources above the 460/1130 mJy threshold described above.

\subsubsection{Large-Scale}\label{large-scale}

As previously mentioned, the variance in the final \ymaps depended on the chosen set of needlets. They controlled how much large-scale contamination, which was mainly dominated by Galactic dust emission, propagated into the final \ymaps. We analyzed the large-scale variations in the \ymaps using the two-point angular correlation function
\begin{equation}
    \xi(\theta) = \langle \rho(\theta + \delta \theta) \rangle
\end{equation}
which is the mean correlation within angular bins of size $\theta + \delta \theta$. The correlation between coordinates $i$ and $j$ is given by
\begin{equation}\label{eq:spear}
    \rho(X_{i},Y_{j}) = \frac{(X_{i}-\mu_{X})(Y_{j}-\mu_{Y})}{\sigma_{X}\sigma_{Y}}
\end{equation}
where X and Y are two maps of the sky, $\mu$ and $\sigma$ represents their means and standard deviations respectively. The two-point correlation function was measured for each all-sky \ymap after applying a set of masks, including the $40\%$ sky coverage galactic mask and the point source mask from \citet{PlanckYMAP}. We also excluded known clusters from the MCXC and PSZ2 catalogs. The reason for applying these masks was to help isolate the general background from small-scale contamination.

    

\begin{figure}[t!]
\centering
     \includegraphics[width=0.49\textwidth]{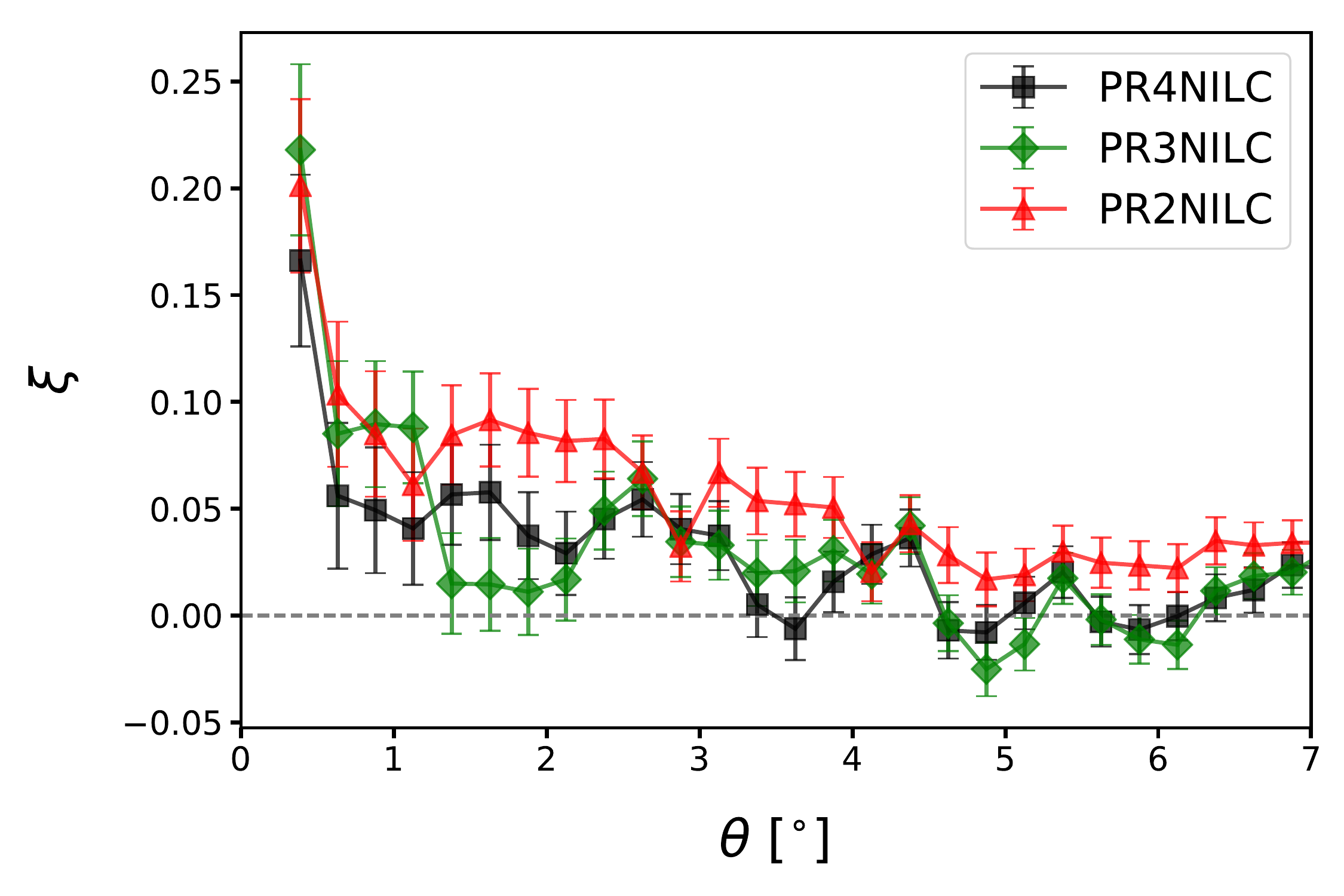}
\par \vspace{1em}
\includegraphics[width=0.49\textwidth]{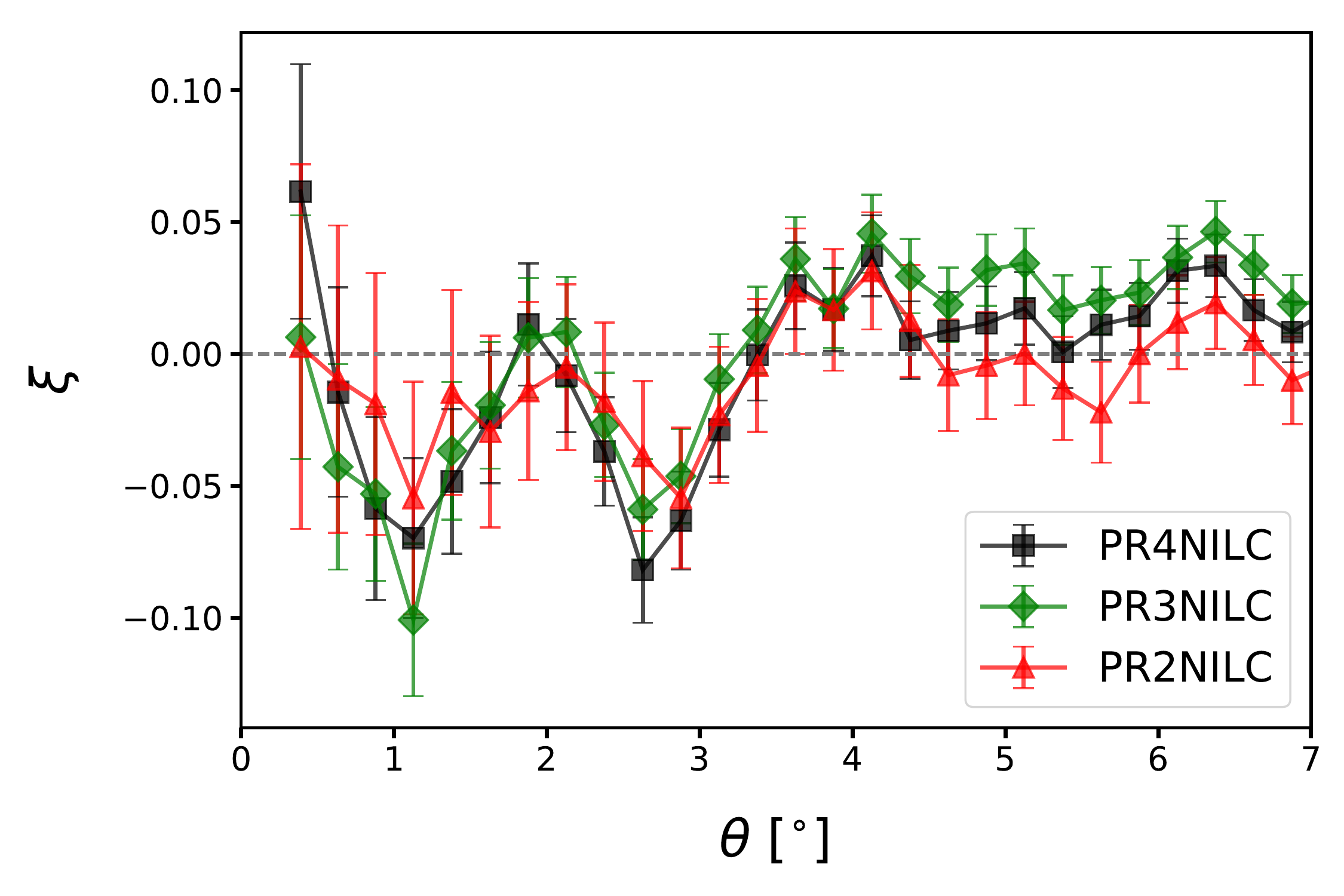}
\vspace{-2em}
\caption{(\textit{top panel}) SZ two-point angular autocorrelation profiles for different \ymaps binned by $\frac{1}{4}^{\circ}$ increments with the first bin omitted. Sources of contamination such as the galactic plane, galaxy clusters, and point sources, were masked for this calculation. Uncertainties were determined in a Monte Carlo manner. PR2NILC showed the strongest background correlations on most angular scales while PR3NILC and PR4NILC appeared similar beyond $2^{\circ}$. (\textit{bottom panel}) Same as the top but represents the angular cross-correlation between the \ymaps and the 857 GHz map which was used as a proxy for the dust emission. Once can see strong anti-correlations on scales $\sim 1^{\circ}$ and $3^{\circ}$.} %
\label{fig:two_point_allsky}%
\end{figure}

The two-point angular autocorrelations of the different \ymaps are shown in the top panel of \autoref{fig:two_point_allsky} where one can see significant correlations on various angular scales for all \ymaps. Most of the correlation power was suppressed beyond 30{\arcmin} (coinciding with the PSF), but significant correlations still existed on larger scales. The apparent peaks and valleys in the correlation functions were likely attributed to the differences in the \Planck data reduction as well as the different NILC methods employed. Furthermore, the PR2NILC profile showed stronger correlations compared to PR3NILC and PR4NILC on nearly all angular scales. Since PR2NILC used different data and window functions compared to PR3NILC and PR4NILC, it was not clear which factor was most responsible for producing the larger correlations. On the other hand, PR3NILC and PR4NILC used identical window functions so the differences in their correlations were purely attributed to the differences in the \Planck data processing. 

On scales $\lesssim 1^{\circ}$, the correlations in PR3NILC were slightly larger than PR4NILC, but the two profiles became consistent $\gtrsim 2^{\circ}$. The observed differences on small scales could be attributed to the different destriping baselines used in cleaning the PR4 and PR3 time-ordered data (see \autoref{sec:discussion}).

The bottom panel of \autoref{fig:two_point_allsky} shows the cross-correlation between the \ymaps and the 857 GHz map, which was used as a proxy for the dust emission. One can see strong anti-correlations between the \ymaps and the dust near $1^{\circ}$ and $3^{\circ}$ for all \ymaps considered. These correlations induced background fluctuations that could affect the extraction of SZ signals. In order to account for strong effects from the dust in our SZ extraction, we excluded regions where the \ymaps were highly correlated with the 857 GHz map. These regions were identified by first calculating correlation maps i.e., multiplying the scaled versions (unit variance and zero mean) of the \ymaps and 857 GHz maps. Next, we smoothed the correlation maps by a 2D Gaussian kernel with FWHM $= 1/2^{\circ}$ as a low-pass filter. Finally, regions bounded by 2$\sigma$ contours were identified as strong dust contamination and excluded in the profile extraction. 

We conclude our discussion of contamination by noting that the stacked SZ profile (see \autoref{sec:results}) remained remarkably unchanged even when these sources of contamination were not excluded. This was reassurance that our final stacked profile was not at the mercy of unidentified contaminants.

\subsection{Uncertainty}\label{sec:uncertainty}
The noise levels in the \ymaps varied across the sky. Here we describe the procedure for evaluating the local background fluctuations around individual objects.

First, we created square images centered on each object in \autoref{tab:sample_info} and made them 20 \Rvir $\times$ 20 \Rvir in size. Then a grid of 36 annuli was placed onto each field, each spaced by 0.25 \Rvir and extending out to 5 \Rvir. They were arranged into a square grid such that the bins inside \Rvir from one annulus did not intersect with the same bins of adjacent annuli. Second, we removed sources of contamination and calculated the mean value of each bin. Third, the mean value of the combined outer 4-5 \Rvir bins was subtracted from the inner annuli as a correction for the local background; this was based on the assumption that the SZ flux beyond 4 \Rvir was negligible. Fourth, we compared alike bins (i.e., those covering the same radial range) only if $>80\%$ of their areas were unaffected by the contamination removal. Finally, the uncertainty was characterized for each field by plotting the relation between the standard deviation of alike bins as a function of annulus area. These data were modeled as a power law, allowing the slope and normalization to vary as free parameters. One would expect the uncertainty for a Gaussian-like background to scale as $A^{-0.5}$ with, $A$, being the uncontaminated area in an annulus; instead, we found slopes closer to $\sim 0.4$.

\subsection{Stacking Methods}
The SZ radial surface brightness profiles were extracted using circular annuli centered on the positions labeled in \autoref{tab:sample_info}. The annuli were scaled to \Rvir and separated by a bin width of 0.25 \Rvir, extending out to 5 \Rvir (similar to the uncertainty analysis). Again, the bins from 4-5 \Rvir were used to determine the local background. Their mean value was subtracted from the rest of the inner bins as a local offset correction. The uncertainty on the local offset was propagated through the stack using the error on this mean value. In addition, the innermost regions were masked by a circle with a 20{\arcmin} radius to avoid expected contamination coming from the member galaxies in each group. 

The surface brightness profiles were stacked by: (1) calculating the weighted average (2) taking the median and bootstrapping to estimate uncertainties. In the first method, bins were weighted by the amount of available area. Since all of the annuli were scaled to \Rvir, the most massive and nearby (i.e., most resolved) objects possessed the lowest error bars. This allowed some objects to carry more weight in the stack. For example, if the area of a bin for one object was twice as large as that of another object, then the uncertainty for the first object would be smaller by roughly a factor of $\sqrt{2}$. On the other hand, the bootstrapping method assigned equal weights to all objects. In \autoref{sec:montecarlo}, we show with Monte Carlo simulations that weighted averaging and bootstrapping yielded similar results. Moving forward, however, most of the discussion will consider the results from the bootstrap method. The uncertainties in bootstrapping were $\sim 30\%$ larger than those in the weighted average estimate, so we adopted the larger uncertainties as a conservative measure. We provide the results of the weighted average stacked profile in \autoref{sec:weighted_avg}. In addition, we tested two sub-samples to reveal any differences/similarities between the 2 most massive and 8 least massive systems.

\subsection{Bias}\label{sec:bias}
The NILC method should suppress the SZ signal due to the well-known ILC bias \citep{Delabrouille09}. Here we describe how we corrected for biases in our sample through Monte Carlo simulations. 
\begin{figure*}[t!]
    \centering
    \includegraphics[width=0.75\textwidth]{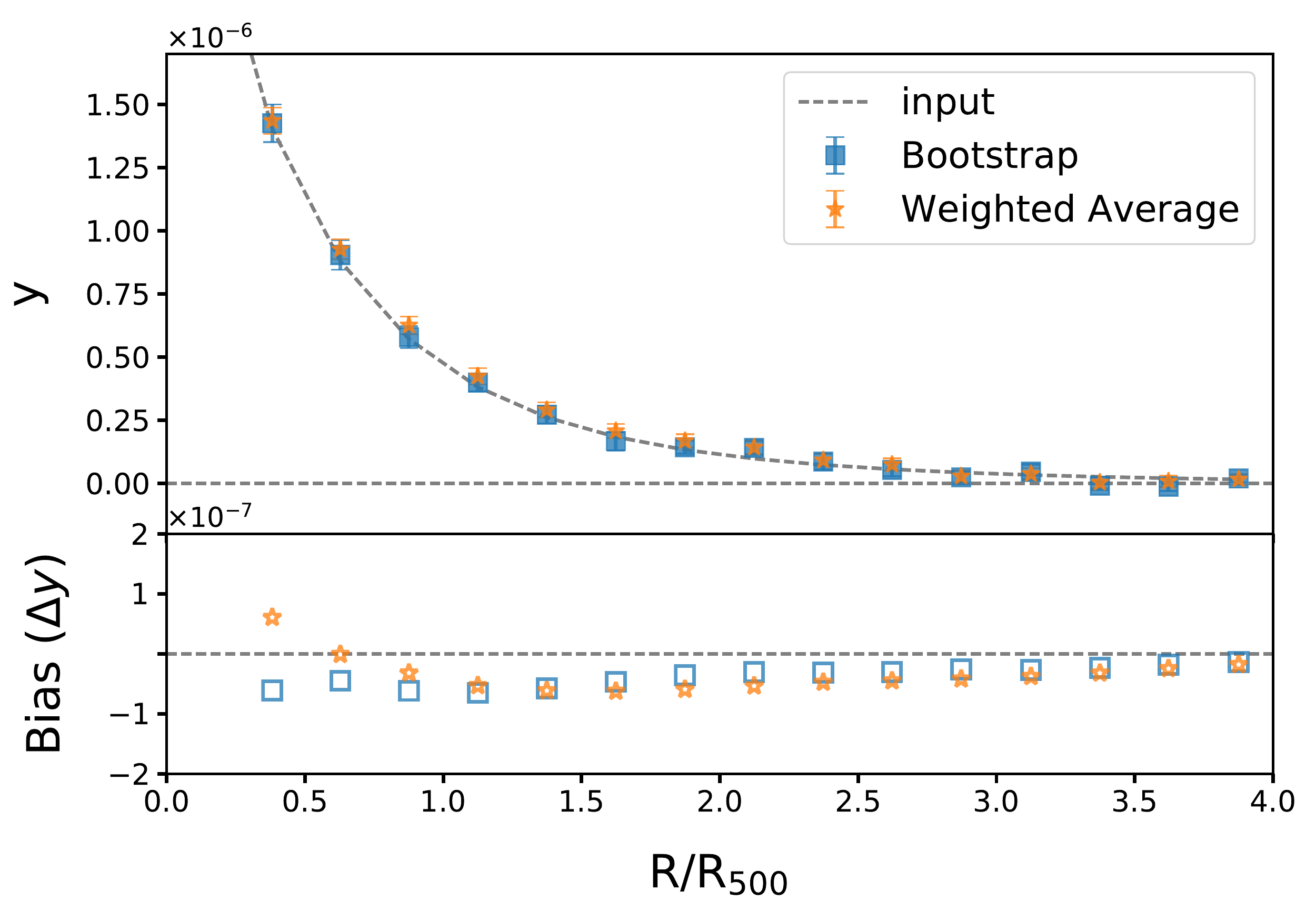} %

    \caption{Results of stacking 100 simulated galaxy group profiles that were injected into the \Planck and \WMAP frequency maps, and recovered in the PR4NILC maps. (\textit{top panel}) The blue squares represent the recovered stacked profile using the bootstrap method, and the orange stars were calculated with a weighted average. Strong sources of contamination were removed before each extraction, and bias corrections were applied from the bottom panel. The dotted line denoted the input signal defined to be the median of all AGN 8.0 profiles from \citetalias{LeBrun15}, each convolved with the PSF. (\textit{bottom panel}) The open markers are the estimated bias values calculated from ``refined'' SZ profiles. They represent the simulated profiles after subtracting off the real SZ portion covering the same region, leaving behind a contamination-free signal. Refined profiles quantified the amount of bias introduced by our NILC algorithm and stacking procedures, which was $\lesssim 5\%$.}%
    \label{fig:simulations}%
\end{figure*}
\begin{figure*}[t!]
\centering
   \includegraphics[width=0.75\textwidth]{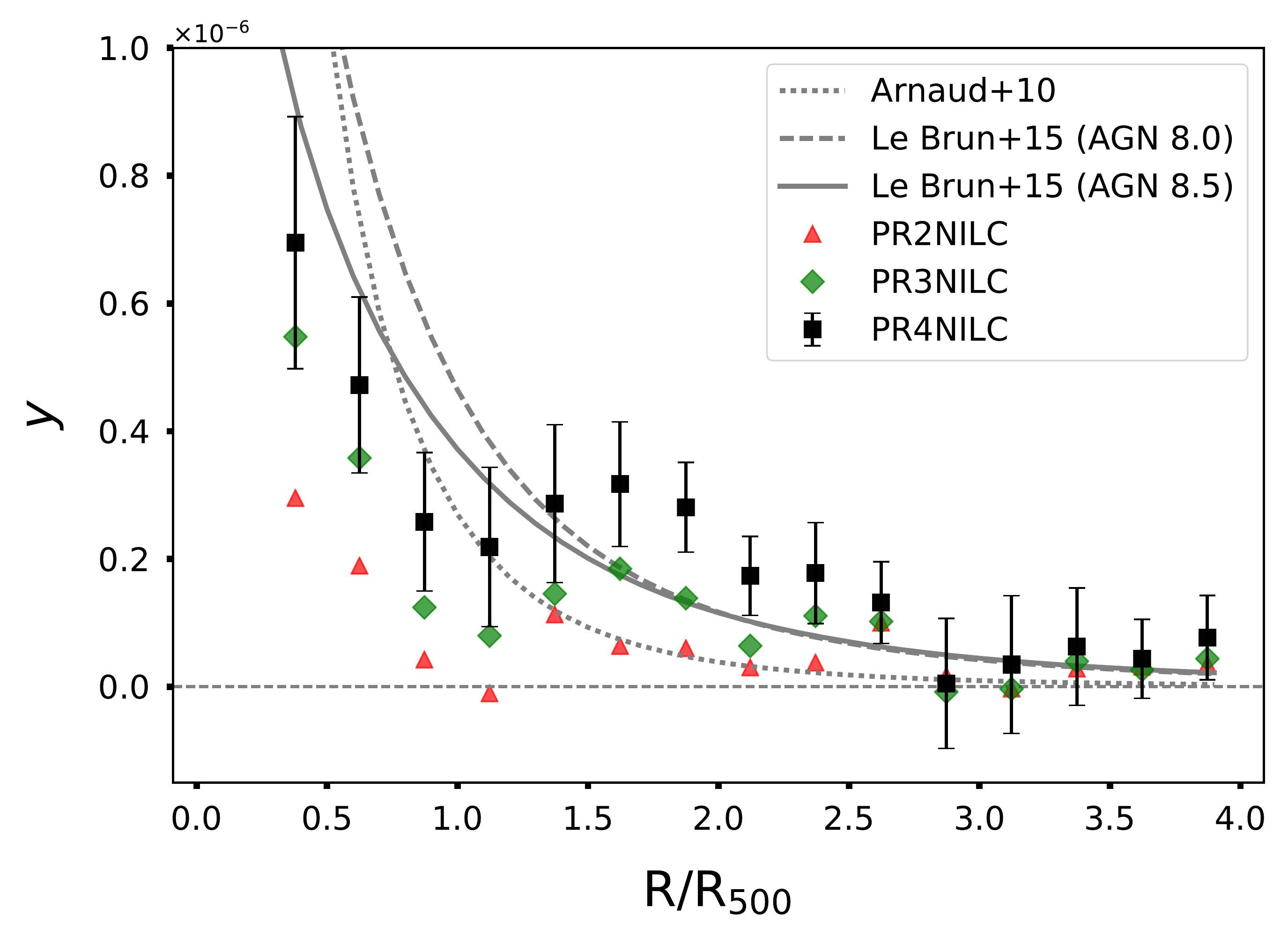}
   
    \caption{The stacked sample of galaxy groups using three different \ymaps. PR2NILC (red triangles) represents on of the public \ymaps while PR3NILC (green diamonds) and PR4NILC (black squares) represent the extractions from our own NILC extraction using later versions of \Planck data with the addition of \WMAP data. The stacked values and their uncertainties were calculated using the bootstrap method. Uncertainties are shown only for PR4NILC for clarity, but the errors for the other profiles are similar. Bias corrections were applied to PR3NILC and PR4NILC data, but this was not possible for PR2NILC data. The dotted gray line is the predicted profiles from \citetalias{Arnaud10}, while the dashed and solid lines are two different AGN feedback models from \citetalias{LeBrun15}.}%
    \label{fig:stack}%
\end{figure*}

\subsubsection{Model Profiles}
The SZ surface brightness ($y$) at some projected radius, $r$, is proportional to the pressure integral along a line-of-sight such that
\begin{equation}
    y(r) = \frac{\sigma_{\mathrm{T}}}{m_{\mathrm{e}}c^{2}} \int_{r}^{R_{\mathrm{max}}} \frac{2P(r')r'}{\sqrt{r^{'2} - r^{2}}}dr'
\end{equation}
where $r'$ is the physical radius, $P(r')$ is the pressure at $r'$ assuming spherical symmetry, $\sigma_{\mathrm{T}}$ is the Thomson cross section, $m_{\mathrm{e}}c^{2}$ is the rest-mass energy of an electron, and $R_{\mathrm{max}}$ is the maximum radius of a halo which we set to 4 \Rvir. 

A common way to model the radial pressure profiles of groups and clusters is with a generalized Navarro Frenk and White (GNFW) model \citep{Nagai07}
\begin{equation}
    \frac{P(r)}{P_{500}} = \frac{P_{0}}{x^{\gamma}(1+x^{\alpha})^{\frac{\beta-\gamma}{\alpha}}}
\end{equation}\label{eq:UPP}
where $x=\frac{c_{500}r}{R_{500}}$. The model is characterized by five parameters: $P_{0}$ is the normalization; $c_{500}$ is a concentration parameter; $\gamma$, $\alpha$, and $\beta$ describe the slopes near the central, middle, and outer regions ($r<<\frac{R_{500}}{c_{500}}$, $r\sim\frac{R_{500}}{c_{500}}$, and $r>>\frac{R_{500}}{c_{500}}$) respectively. Many of these parameters have been constrained through observations of massive clusters provided by the seminal works of \citet[][hereafter \citetalias{Arnaud10}]{Arnaud10} and \citet{PlanckClusterProfile}. Observational data near the cluster outskirts, however, are generally too poor to fit for the outer slope; thus $\beta$ is oftentimes fixed using results from numerical simulations.

We considered three different GNFW models to describe the stacked SZ profile. First was the predicted profile from \citetalias{Arnaud10}. This profile was derived from an X-ray sample of galaxy clusters out to \Rvir. One might expect galaxy groups to exhibit this shape if they were simply scaled-down versions of galaxy clusters. The other two models were derived from the simulations of \citetalias{LeBrun15} with different AGN feedback prescriptions, i.e., AGN 8.0 and AGN 8.5. These feedback prescriptions allowed the central black hole to accrete gas and build energy until it could heat the ambient gas by a pre-defined quantity, $\Delta T_{\mathrm{heat}}$. This value was set to $10^{8}$ K and $10^{8.5}$ K for the AGN 8.0 and AGN 8.5 models respectively. Once the appropriate amount of energy was gained by the black hole, then $1.5 \%$ of the rest-mass energy of the accreted gas was released as feedback. Since the AGN 8.5 prescription required more gas to be accreted before feedback could take place, it resulted in more bursty outflows relative to the AGN 8.0 model.

\subsubsection{Monte Carlo Simulations} \label{sec:montecarlo}
In order to understand the biases in our sample, we injected 100 simulated objects into the PR4, and \WMAP frequency maps. The simulated sample was drawn from the mass and redshift properties from \autoref{tab:sample_info}. Each object in the sample was simulated 10 times and injected into 10 random locations on the sky (100 total) at galactic latitudes $|b|>15^{\circ}$. We adopted the AGN 8.0 model from \citetalias{LeBrun15} as the fiducial model for the simulated SZ profiles (we refer to this as the ``fiducial model'' herein). The fiducial model for each object was added into the frequency maps using \autoref{eq:sz_temp} after convolving the signals with the PSF (assumed to be a 2D Gaussian with a FWHM equal to the instrumental resolution).

The bias was calculated by subtracting the real \ymap (i.e., PR4NILC with no injected sources) from the simulated \ymap. This resulted in a ``refined'' \ymap that was free of contamination. The refined profiles were recovered in the same way as the real groups, i.e., annuli bins were separated by 0.25 \Rvir, the outer annuli from 4-5 \Rvir were subtracted from each field to correct for local background offsets, and the central 20{\arcmin} regions were excluded. One can then measure the bias by comparing the refined profile with the true input. The true input was considered to be the median of all fiducial profiles generated for each object convolved by the PSF. The difference between the refined signal and the input values are shown in the bottom panel of \autoref{fig:simulations}. 

The bias profiles were slightly different among extraction methods. Most notably, the inner bins contained positive bias in the weighted average method while it was entirely negative for the bootstrap technique. While one would expect the ILC bias to always be negative, the positive bias in the weighted average was attributed to the unequal contributions from the most resolved objects. Indeed, the two most massive objects in the sample happened to be the most extended on the sky. Therefore, their estimated errors were lower and more weight was given to them in the stack (see \autoref{sec:uncertainty}). Conversely, the bootstrap method assumed equal weighting for each object. 

In the top panels of \autoref{fig:simulations} we show the recovered simulated profiles using our extraction method. These were different than the refined profiles because they were corrected for the bias and also susceptible to residual contamination. This experiment demonstrated our ability to recover the stack of SZ profiles. Specifically, we measured the reduced $\chi^{2}$ ($\chi^{2}_{\mathrm{red}}$) = 0.78 (1.03) for the bootstrap (weighted average) stack for 15 degrees of freedom (DOF).

\section{Results}\label{sec:results}

This section compares different model predictions with the stacked surface brightness and cumulative profiles. First we compare the data and models at face-value, and then we perform a least-squares fit. 

\subsection{Surface Brightness Profile}
The stacked (bootstrapped) signal of galaxy groups from different \ymaps{} is shown in \autoref{fig:stack} along with three different GNFW model predictions for our sample. (The individual profiles are provided in \autoref{fig:individual} in the appendix.) At face value, none of the model predictions appeared consistent with the stacked profiles from PR2NILC and PR3NILC. The profile from PR4NILC, however, was measured using better data, and this was in agreement with the the AGN 8.5 model prediction; the $\chi^{2}_{\mathrm{red}}$ value using the bootstrap (weighted average) method was 1.23 (1.15) for 15 DOF. 

Next, we fit each data set using the three model profiles described above, specifically fixing their shapes and allowing the normalizations to vary as free parameters. The best-fit normalizations using PR2NILC yielded values 18\%-25\% of the original models; those for PR3NILC data (bias corrected) were 41\%-57\%; and PR4NILC (bias corrected) returned 60\%-92\%. The AGN 8.5 profile fit (the flattest of the three models) to PR4NILC data yielded the largest normalization of $92\%$ with respect to the original prediction. It also gave the lowest $\chi^{2}_{\mathrm{red}}$ value of 1.28 (14 DOF) compared to the other model fits. 

Out of curiosity, we also extracted the stacked signal from the public MILCA \ymap \citep{PlanckYMAP} and found it was nearly identical to that from PR2NILC. This result suggests the PR2 data processing could be suppressing the SZ signal and, in some cases, this may be even more important than the chosen ILC method. (See \autoref{sec:data release} for the discussion on different data release versions.) We do not further report on the MILCA \ymap and focus on only the NILC method.

\subsection{Extended SZ Profile}
The stacked SZ profile exhibited a ``bump'' feature around 1-3 \Rvir as seen in \autoref{fig:stack}. This feature appeared both in the weighted average and the bootstrap stacks, so the feature was not due to the stacking procedure. In addition, the feature was not dominated by a few objects as the median and mean stacking yielded similar results. The significance of the bump was determined by fitting the fiducial model to the data but excluding those within the arbitrary region of 1.25-3 \Rvir. This was done both for the simulated and real data sets.

\begin{figure}[htb]
\centering
   \includegraphics[width=0.45\textwidth]{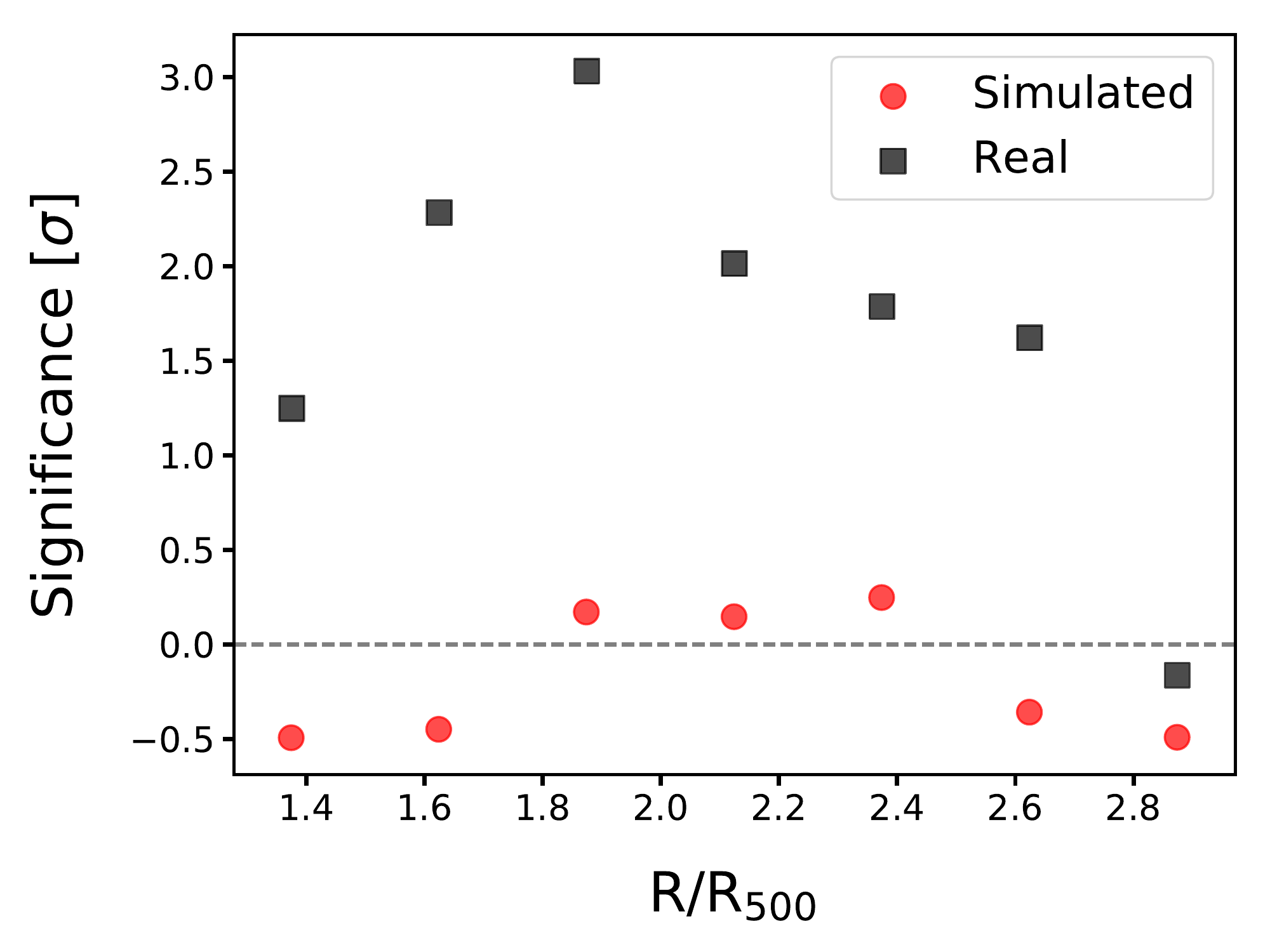}
    \caption{Significance profiles when trying to measure a bump feature both in the simulated and real data sets. First, AGN 8.0 profiles were fit to the data outside of the bump region (1.3 - 2.9 \Rvir). Then residuals were calculated between the best-fit model and the data over the bump region. The significance was calculated by simply dividing each residual by the uncertainty of the measurements.}%
    \label{fig:bump}%
\end{figure}

\begin{figure}[b!]
\centering
   \includegraphics[width=0.45\textwidth]{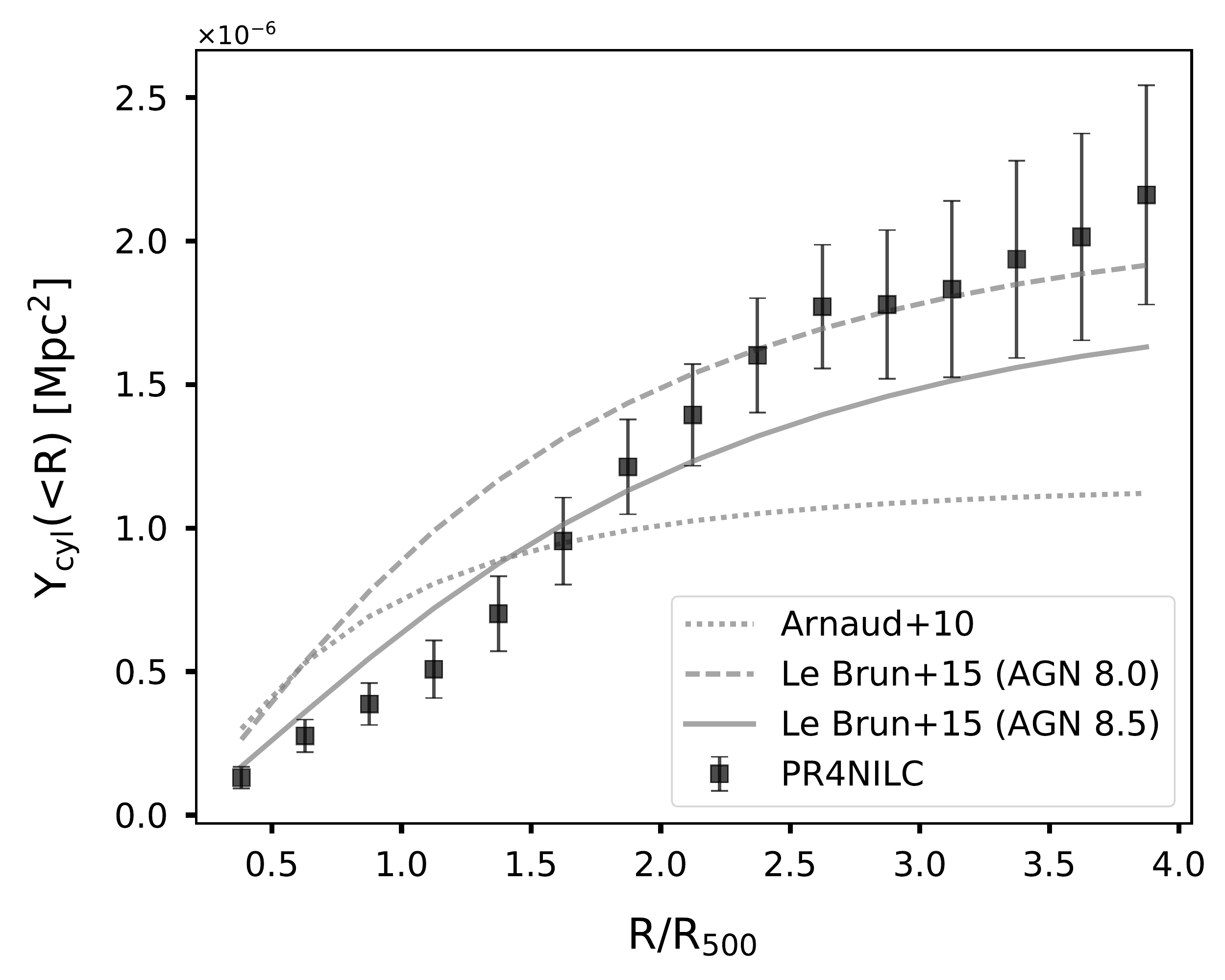}
    \caption{Cumulative cylindrically integrated profiles. The black data represent the cumulative $Y_{\mathrm{cyl}}$ for the PR4NILC data while the different lines denote the three model predictions considered in this study.}%
    \label{fig:cumulative}%
\end{figure}

The residuals between the best-fit model and the data in the bump region are shown in \autoref{fig:bump}. There were $>2\sigma$ deviations between the model and real data around 1.6-2.2 \Rvir, with a peak significance of $3.0\sigma$ near 2 \Rvir. We note these calculations were mildly model dependent. Specifically, the maximum significance around 2 \Rvir was 3.0, 3.0, and 2.6 $\sigma$ when fitting the \citetalias{Arnaud10}, AGN 8.0, and AGN 8.5 models respectively. On the other hand, the results from the simulated data did not yield significant deviations over the bump region. These results suggested the bump was unlikely an artificial feature from the NILC method.

\subsection{Cumulative Signal}
The cylindrical integrated SZ quantity within a projected radius, $r$, is given by
\begin{equation}
    Y_{\mathrm{cyl}}(r) = \sum_{i; r_{i}<r} 2\pi r_{i} y(r_{i}) \Delta r
\end{equation}
where $i$ is the index of each radial bin, $y$ is the SZ signal, and $\Delta r$ = 0.14 Mpc (equal to 0.25 $\times$ the median \Rvir of the sample). It measures the product of the gas mass and mass-weighted temperature enclosed within a cylinder along the line-of-sight. At large enough radii $Y_{\mathrm{cyl}}$ captures the total SZ flux of the system which is a proxy for the total thermal energy of the halo plasma. 

The cumulative $Y_{\mathrm{cyl}}$ is shown in \autoref{fig:cumulative} along with the three different model predictions. None of the models were consistent with the inner parts of the data, however, the cumulative signal converged with the AGN 8.0 prediction from \citetalias{LeBrun15} near $\sim 2.5$ \Rvir. This places a crucial constraint on the amount of thermal energy bound inside galaxy groups that must be matched by numerical simulations.

\section{Discussion}\label{sec:discussion}
This section presents our interpretation of the results. We start by discussing the discrepancies among different \ymaps. Then we estimate the baryon fraction and compare our value with previous studies. Finally, we discuss the nature of the extended SZ profile.

\subsection{Data Release Versions}
\label{sec:data release}
In \autoref{fig:stack} one can see the signal from PR3NILC appeared larger than PR2NILC for radii $\lesssim 2.5$ \Rvir. At the same time, the signal from PR4NILC superseded that of PR3NILC. This last statement points to the differences in the data processing between PR3 and PR4 data. 

While there were many improvements from PR3 to PR4, the best explanation for the observed differences was due to the effects of ``striping''. Turning time-ordered data from multiple scans of the sky can introduce residuals along the scanning pattern of the satellite. Thermal instabilities, far sidelobes, and elliptical beam shapes are known to produce stripes in the all-sky frequency maps. Multiple algorithms have been developed to reduce these features \citep{Poutenan04,Keihanan04} and have made major improvements in recent years.

\begin{figure}[htbp]
\centering
   \includegraphics[width=0.45\textwidth]{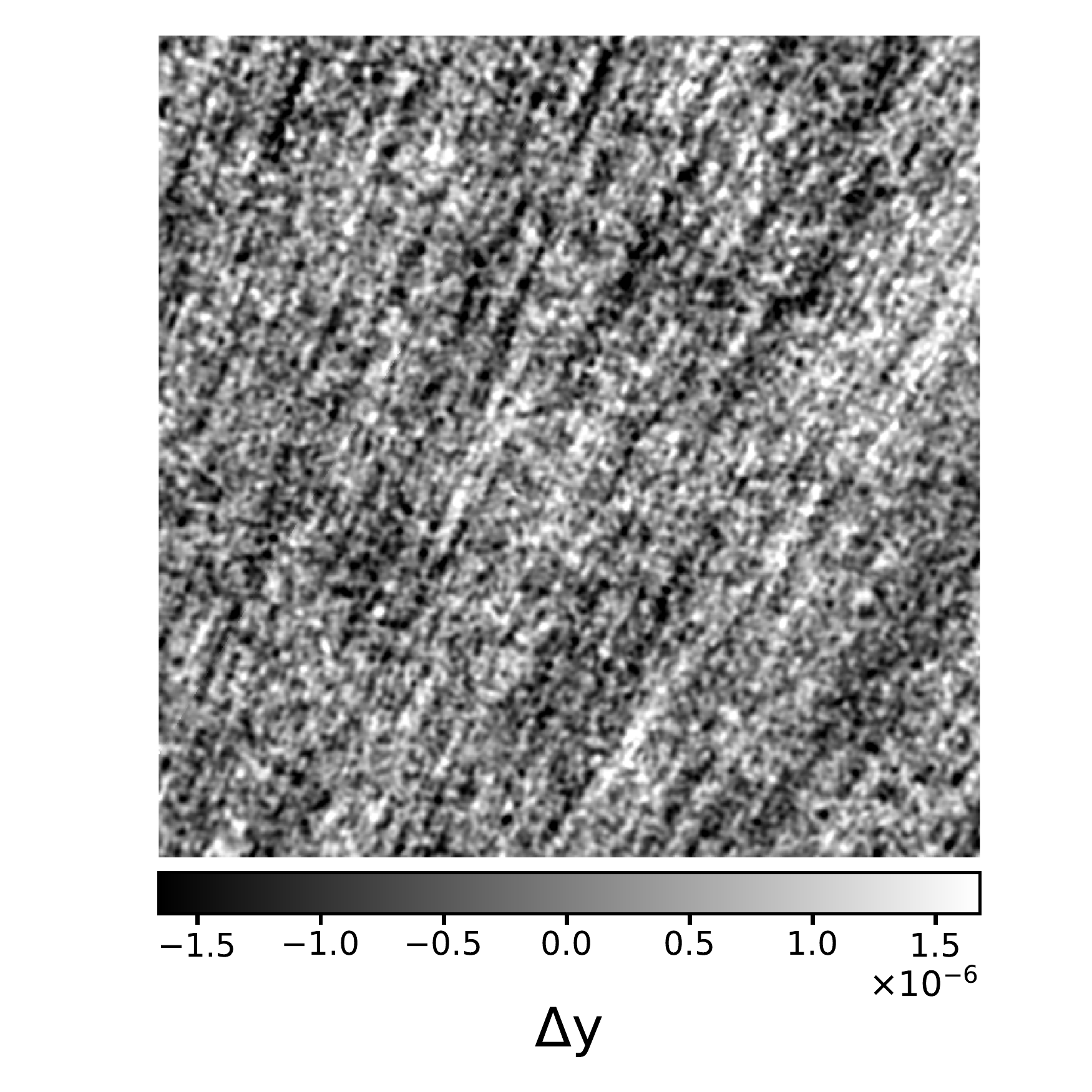}
    \caption{A randomly selected patch showing the difference between PR4NILC and PR3NILC. The region is centered at the coordinates ($l=161.431^{\circ}$, $b=84.516^{\circ}$) with a field-of-view $22^{\circ}\times 22^{\circ}$. There are clear striping features that coincide with the \Planck scanning pattern. Since the NILC method was the same for PR3NILC and PR4NILC, the difference in the \ymaps is purely due to data processing.}%
    \label{fig:PR4-PR3}%
\end{figure}

The PR4 destriping technique used a baseline of 167 ms which was much shorter than the baselines used on PR3 data \citep{PlanckPR4}. The PR4 baseline corresponded to roughly $1^{\circ}$ on the sky, which may be why the PR4NILC correlation function was smaller on sub-degree scales relative to PR3NILC. It may also be why the SZ profile from PR3NILC appeared suppressed relative to PR4NILC in \autoref{fig:stack}. The difference map between PR4NILC and PR3NILC is shown in \autoref{fig:PR4-PR3}, serving as a visualization of the striping features.

\subsection{Baryon Fraction}
We calculated the amount of hot gas in the halo using the best-fit pressure profiles to all of SZ data (bump region included). The SZ effect measures the pressure of the electrons $P_{\mathrm{e}}(r) = n_{\mathrm{e}}(r)k_{\mathrm{B}}T_{\mathrm{e}}(r)$ where $n_{\mathrm{e}}$ is the number density of electrons. We solved for $n_{\mathrm{e}}(r)$ using our best-fit pressure profile (AGN 8.5 model) to the SZ data and assuming an exponentially decaying temperature profile from \citet{Sun09}. Furthermore, we used their $M_{500}-T_{500}$ relation and result of $\frac{T_{500}}{T_{2500}} = 0.89$ and adopted a metallicity of $Z = 0.3 ~Z_{\odot}$ to derive the gas mass density $\rho_{\mathrm{gas}}(r)$. Integrating this quantity out to \Rvir yielded $f_{\mathrm{gas},500} = 0.038\pm 0.007$, that is the gas mass fraction relative to the total mass within \Rvir.

The baryon fraction within \Rvir, $f_{\mathrm{b},500}$, was calculated by adding the stellar mass to the gas mass. The stellar mass was estimated using the $K_{s}$ luminosity presented in the \citetalias{KTully17} catalog and assuming a mass-to-light ratio $\sim 0.8$ in solar units \citep{Bell03}. The stellar mass fraction within \Rvir was $f_{*,500} \approx 0.018$, yielding a baryon fraction of $f_{\mathrm{b},500} \approx 0.056$, which is much smaller than the cosmic value of 0.16 \citep{PlanckCosmoValues}. 

Additional systematic uncertainties are expected when assuming a pressure profile. In fact, all of the stacked data within \Rvir appeared to lie systematically below the best-fit profiles, while those in the bump region resided above the fit. Thus, it is possible the baryon fraction within \Rvir was overestimated by the best-fit models.

Observational studies have showed $f_{\mathrm{gas},500}$ tends to decrease with decreasing halo mass \citep{Dai10, Lagana13}. For galaxy groups, there is a general consensus that $f_{\mathrm{gas},500}$ is $\lesssim 10\%$. For example, \citet{Tholken16} measured $f_{\mathrm{gas},500} \approx 9\%$ for the galaxy group UGC 03957 using mosaic images from Suzaku. Similarly, \citet{Sarkar20} measured $f_{\mathrm{gas},500} \approx 7\%$ using joint Suzaku and Chandra observations of the galaxy group MKW4. These results, along with those presented in this study, suggest galaxy groups possess $f_{\mathrm{b},500}\sim$ 1/2 the cosmic value. 

Our value of the gas fraction was roughly half compared those reported in the literature. One possible explanation for this discrepancy could be because our sample of galaxy groups was not X-ray selected. X-ray detected groups and clusters tend to be biased toward more virialized systems. While massive galaxy clusters almost always appear relaxed, galaxy groups likely exist in a wide range of dynamical states. Furthermore, it is possible that not all of the gas in galaxy groups has been heated to the virial temperature. If the true temperature of the gas was lower than what was assumed in our models then the inferred gas density would be closer to the estimated baryon fractions determined by X-ray observations.

\subsection{Bump Feature}

Here we consider the possible explanations for an increase in the SZ signal beyond the virial radius. One might argue there should be projection effects from neighboring halos known as the two-halo contribution \citep{Vikram17,Hill18}. This term was calculated as described in \citet{Pandey19} assuming the \citetalias{LeBrun15} pressure profiles. The two-halo contribution was found to be negligible ($y \sim 10^{-8}$ within 4 \Rvir), which was mainly because of the low-redshift nature of the sample. A bump feature may also arise from miscentering during the SZ extraction. We tested this by purposefully misaligning the extraction annuli on the simulated data set, but the results did not show significant increases near the bump region. Another possibility is the bump could be attributed to the anit-correlations between the dust and \ymaps near $1^{\circ}$. Rather than describing the feature as an excess of SZ signal near $2.5$ \Rvir, one could argue for the presence of a ``valley'' near $1$ \Rvir instead. It is possible the anti-correlations between the \ymaps and the dust may have caused a suppressed SZ signal. Nevertheless, we did not see strong evidence for systematic bumps or valleys in the Monte Carlo simulations (\autoref{sec:montecarlo}).

One might also wonder if AGN outbursts could sweep gas out to very large radii. In other words, the bump feature seen in these results could be a giant pile-up of material that was expelled by the AGN. Below we give an order-of-magnitude argument as to why this is energetically possible.

In this experiment, we applied a model that could capture the bump feature in the SZ data. This was done by over-fitting the data with a fifth-degree polynomial pressure model. Specifically, we assumed an extrapolated temperature profile from \citet{Sun09} to get an estimate of $n_{\mathrm{e}}$, which was then converted into a gas mass density, $\rho_{\mathrm{gas}}$. We then calculated a gas mass reservoir of log$_{10}(M_{\mathrm{gas}} [M_{\odot}])\approx 13.2$ over the volume covered by the bump region (1.25 - 3 \Rvir). For comparison, the fiducial AGN 8.0 model predicted a gas mass of log$_{10}(M_{\mathrm{gas}} [M_{\odot}])\approx 13.0$.

Driving a significant amount of gas to large radii changes its binding energy. The binding energy of gas within some physical radius, $r$, is given by
\begin{equation}
    E_{\mathrm{bind}} = 4\pi G \int_{0}^{r}M(<r^{\prime})\rho_{gas}(r^{\prime})r^{\prime} dr^{\prime}
\end{equation}
where $G$ is the gravitational constant, $M(<r)$ is the enclosed mass assumed to be dominated by the dark matter. The dark matter density profile was estimated using an NFW profile with a concentration parameter $c_{500} = 3.2$ \citep[e.g.,][]{Holland15}. The two models of the gas densities (fiducial vs. polynomial) returned similar binding energies of the gas $\sim 10^{62}$ ergs at 3 \Rvir. This was not surprising given the results in \autoref{fig:cumulative}. At most, the binding energies differed by a few $10^{61}$ ergs which is on the order of the mechanical feedback energy from radio galaxies in galaxy groups \citep{Giodini10}. Thus, it is energetically possible for AGN feedback to cause the bump seen in our data. 

In addition, this density profile (polynomial fit) would require galaxy groups to be out of hydrostatic equilibrium near the outskirts. Our order-of-magnitude calculation suggested the gravitational pressure would be $\sim 2\times$ larger compared the thermal gas pressure around 1.5 \Rvir. In fact, this is in agreement with the simulation results from \citet{Burns10} that showed galaxy clusters are likely not in hydrostatic equilibrium near the outskirts. These authors demonstrated that turbulence and bulk radial motions could contribute half of the total pressure support needed to balance gravity at large distances. 

Instead of piling gas at large distances, it is also possible the extended gas was heated through accretion shocks. Accretion shocks occur when infalling gas encounters extreme changes in density and/or temperature in the gaseous halo. There are two types of accretion shocks: external and internal. External shocks are caused by pristine gas falling onto the halo for the very first time. In general, these are strong shocks with high mach numbers \citep{Ryu03} and are believed to occur at distances of several \Rvir in galaxy clusters \citep{Hurier19,Zhang20,Baxter21}. On the other hand, internal shocks consist of gas that has already passed through the external shock. They are caused by gas falling onto substructure within the halo, subclump mergers, and supersonic flows inside substructures; they also dissipate more thermal energy than external mergers \citep{Ryu03,Pfrommer06, Vazza09}.

Here we present tentative observational evidence of internal shocks in galaxy clusters. At least one SZ study by \citet{Adam18} has reported evidence of substructure in individual clusters. These authors applied structure filters to their data and were able to detect pressure ridges in some systems out to hundreds of kpc. We also noticed plateau-like features in the SZ profiles reported by the Planck Collaboration. Specifically, there was large residuals from the GNFW model fit around 1-3 \Rvir for a stacked sample of galaxy clusters \citep{PlanckClusterProfile}. A similar feature was also apparent in the Virgo cluster where the SZ profile flattened outside the virial radius \citep{PlanckVirgo}. The Coma cluster \citep{PlanckComa} also appeared to have an elevated SZ signal beyond \Rvir compared to the \citetalias{Arnaud10} prediction. One can also see bump-like deviations in the azimuthal profiles of Abell 2319 by \citet{Hurier19}. It is possible the bump feature seen around 1-3 \Rvir in our stacked sample of galaxy groups is similar to those seen in galaxy clusters. In many of these studies, however, internal accretion shocks were not considered. 

Internal shocks are expected to be seen starting near $\sim 0.7$ Mpc  \citep[$\approx$ 1.3 \Rvir for this sample;][]{Ryu03} and extending out to the external shock boundary at $\sim$ 5-6 \Rvir \citep{Baxter21}. It was unlikely for us to detect an external shock since we subtracted off the 4-5 \Rvir regions as local background corrections. The most probable explanation for the bump feature in our stacked data are the manifestations of internal accretion shocks at various radii along a line-of-sight.
\section{Summary}\label{sec:summary}

This work presented the resolved SZ profile for a stacked sample of 10 galaxy groups with a median mass of log$_{10}(M_{500} [M_{\odot}]) = 13.7$. The stacked signal was extracted using the publicly available \ymap as well as our own \ymaps constructed from newer versions of $Planck$ data with the inclusion of \WMAP data. For each field, we excluded the strongest sources of contamination to obtain the most accurate measurements. These signals were stacked using a weighted average and bootstrap method, both of which produced consistent results. Finally, we demonstrated our ability to accurately recover a stacked sample of galaxy groups in our own \ymaps through Monte Carlo simulations. The main conclusions from our study are:
\begin{itemize}

    \item There have been significant improvements to the \Planck channel maps in recent years. We utilized the latest and best version of \Planck data to construct a superior \ymap, PR4NILC, compared to the public \ymap, P2NILC. The PR4NILC data yielded a significant SZ detection out to $\sim$ 3 \Rvir and was in agreement with the AGN 8.5 model prediction from \citetalias{LeBrun15}. Conversely, the PR2NILC data did not show a convincing signal beyond \Rvir and did not agree with any of the model predictions considered.
    
    \item We estimated a gas fraction of $f_{gas,500} = 0.038 \pm 0.007$ and a baryon fraction $f_{b,500} \approx 0.056$ using our best-fit pressure model and assuming a temperature profile for galaxy groups \citep{Sun09}. This was slightly smaller than the values reported by X-ray studies of individual galaxy groups. Since our sample was not X-ray selected, these groups may not have yet heated their baryons to the virial temperature. Our estimated baryon fraction was $\lesssim 1/2$ of the predicted cosmic value (0.16) suggesting AGN feedback could be effective at removing gas from galaxy groups.
    
    \item We detected a $3 \sigma$ bump feature near 2 \Rvir in our stacked SZ profile. This feature was most likely a detection of internal accretion shocks at different radii. We found similar features throughout the literature of galaxy clusters, which further supported this claim.
\end{itemize}

This work has unveiled key discrepancies in extracting weak SZ signals from different versions of \Planck data, and this must be addressed in similar future studies. Future SZ experiments, such as CMB-S4, will have the ability to study individual galaxy groups with high resolution and sensitivity and can use our results as a baseline.

We thank Shivam Pandey at the University of Pennsylvania for calculations of the two-halo contribution specific to our sample. We are grateful for support from NASA through the Astrophysics Data Analysis Program, awards AWD012791 and NNX15AM93G. The work is based on observations obtained with \Planck, http://www.esa.int/Planck, an ESA science mission with instruments and contributions directly funded by ESA Member States, NASA, and Canada. This research made use of the High Energy Astrophysics Science Archive Research Center (HEASARC) and the NASA/IPAC Extragalactic Database (NED), which are supported by NASA.

\newpage
\bibliography{bib.bib}
\appendix
\section{Individual Surface Brightness Profiles}
\begin{figure}[t!]
\centering
    \includegraphics[width=0.95\textwidth]{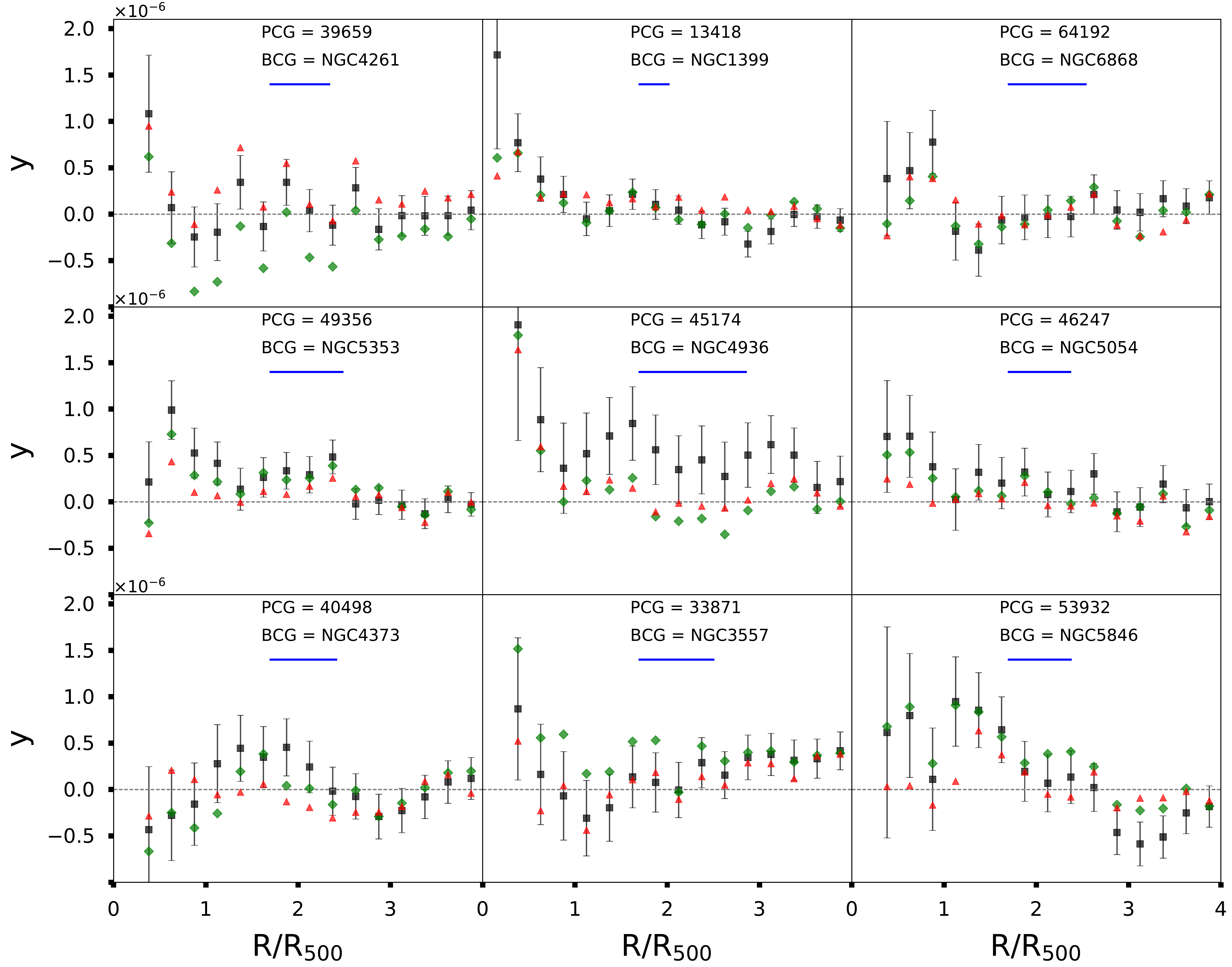}%
    
    \includegraphics[width=0.37\textwidth]{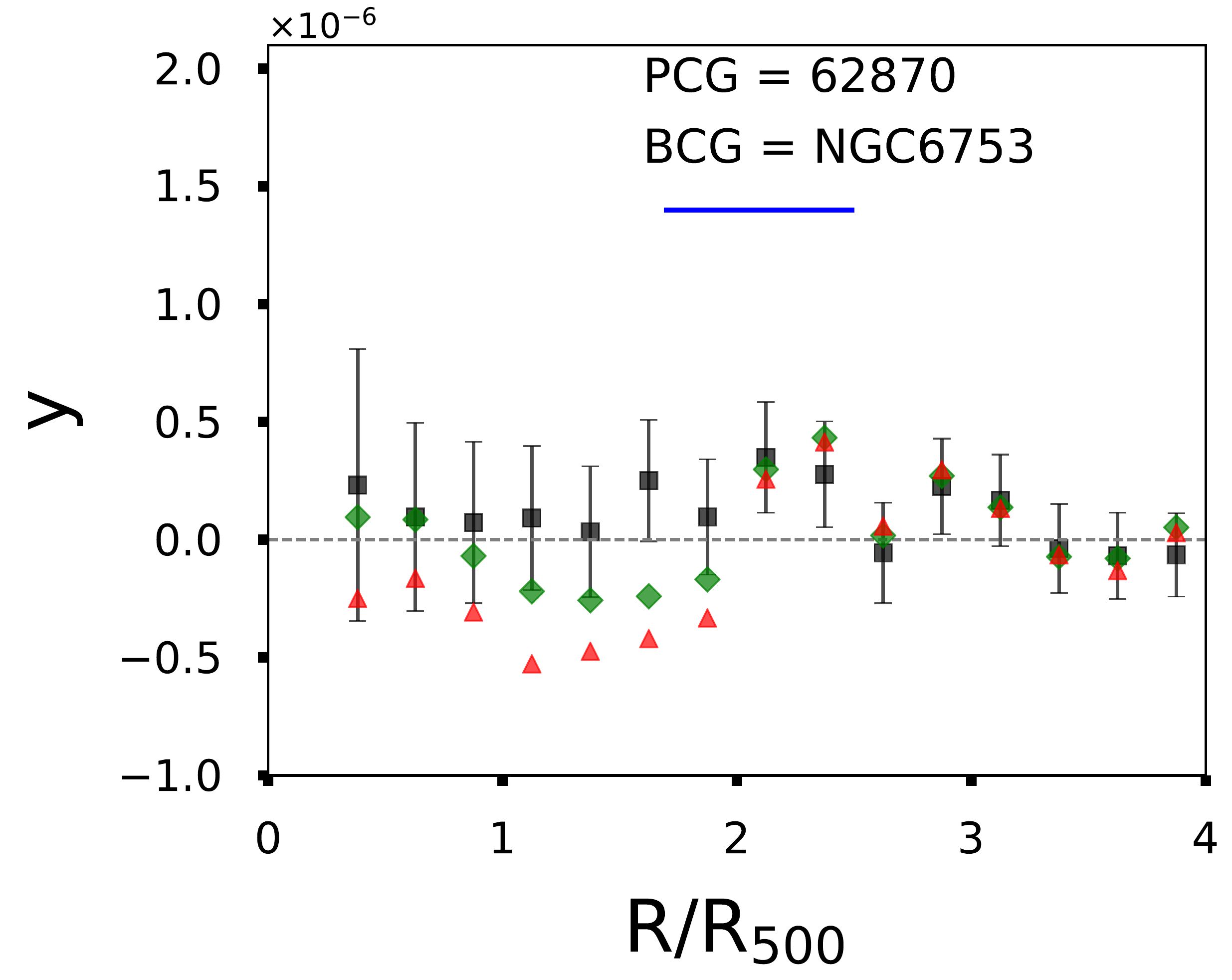}
    
    \caption{Individual SZ surface brightness profiles. Red triangles, green diamonds, and black squares represent the profiles from PR2NILC, PR3NILC, and PR4NILC respectively. Uncertainties are shown only for PR4NILC and were estimated locally depending on the size and location of the extraction annulus (see \autoref{sec:uncertainty}). The PCG identification represents the objects in \autoref{tab:sample_info}; the corresponding BCG IDs are provided as well. Blue lines indicate the $1^{\circ}$ scale for each object.}%
    \label{fig:individual}%
\end{figure}
\newpage
\section{Weighted Average Stacked Profile}
\label{sec:weighted_avg}

Stacking signals together is only acceptable when the spread about the mean is modest, and so one must search for possible for outliers that may be skewing the resulting stack. For the sample in this work, the two most massive systems were the closest in distance. One might expect the combination of their high masses and close distances to yield high S/N values, which could potentially bias the weighted stack. We tested this possibility by splitting our galaxy group sample into
low-mass (8 least massive) and high-mass (2 most massive) sub-samples. (We did not consider the bootstrapping method for the sub-samples due to insufficient sample size of the high-mass objects.)

\begin{figure}[h!]
\centering
    \includegraphics[width=0.75\textwidth]{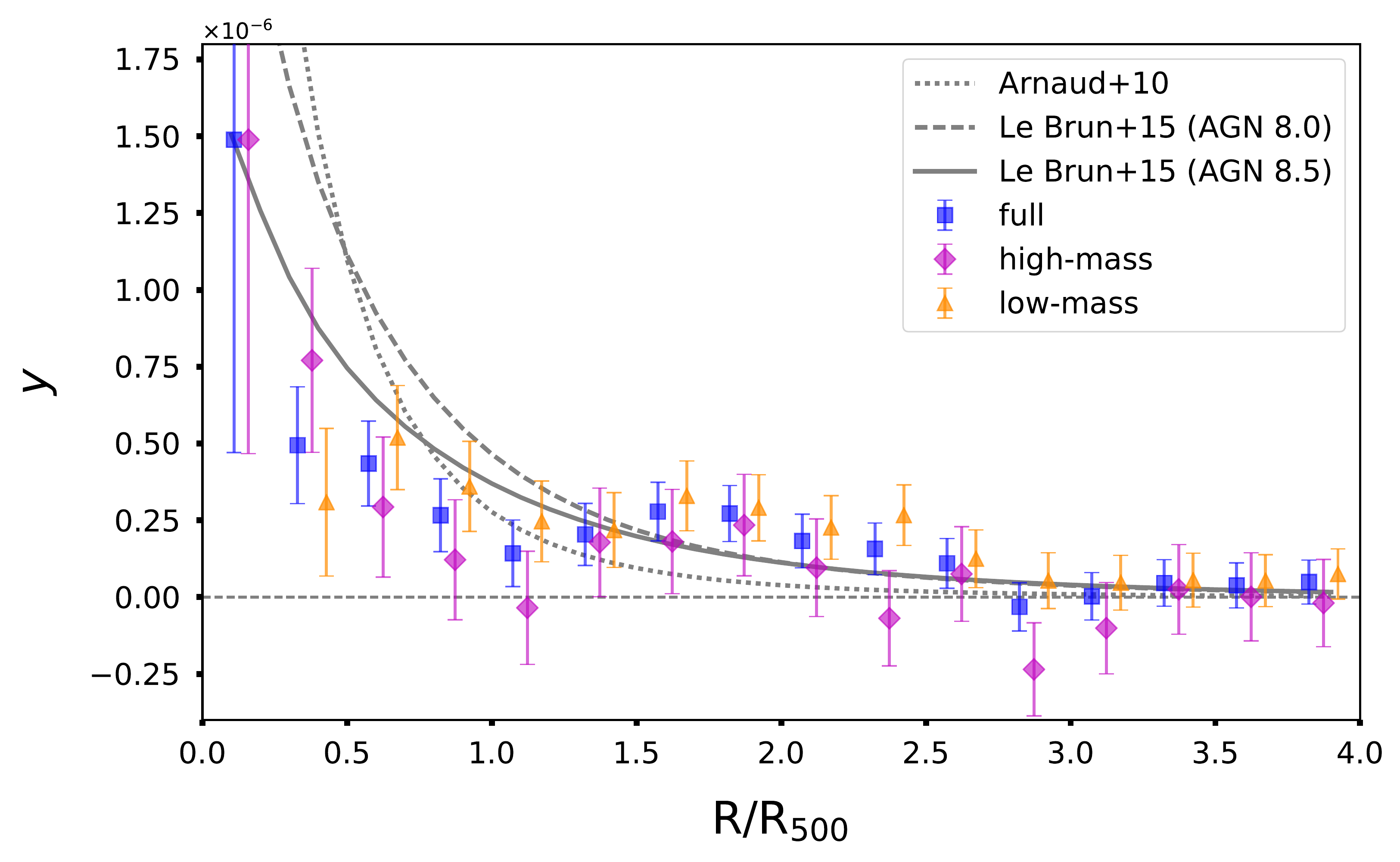}%

    \caption{Weighted average stacked profiles using the full (blue squares), low-mass (orange triangles), and high-mass (purple diamonds) samples. The data are slightly offset on the horizontal axis for clarity; the blue and orange data are shifted to the left and the right of each purple diamond respectively. The weighted stack was calculated using the empirical variance estimated for each local field. The low-mass sample does not show a value for the innermost bin because it was smaller than the 20' exclusion radius (i.e., 0.25 R$_{500} < 20^{\prime}$).}%
    \label{fig:weighted_avg}%
\end{figure}

The stacked profile from the full, low-mass, and high-mass samples are shown in \autoref{fig:weighted_avg}. Tenuous evidence points to a steeper profile for the high-mass sample compared to the low-mass sample within R$<$\Rvir; however, the data were mostly consistent within the uncertainties. In addition, both sub-samples exhibited a bump feature near 1.75 \Rvir, meaning the bump was not dominated by only low- or high-mass systems.

\end{document}